\documentclass[prb,twocolumn,amsmath,amssymb,eqsecnum]{revtex4-1} 

\usepackage{feynmp}
\usepackage{amsmath}
\usepackage{graphicx}
\usepackage{rotating}
\usepackage{multirow}
\usepackage{latexsym}
\usepackage{textcomp}
\usepackage{verbatim}
\usepackage{color}
\usepackage{pict2e}
\usepackage{bm}
\usepackage{subfigure}
\usepackage{everysel}
\usepackage{keyval}
\usepackage{ragged2e}
\usepackage{dsfont}
\usepackage{amssymb}
\usepackage{enumitem}
\usepackage{pstricks}
\usepackage{mathtools}
\usepackage{hyperref}

\newcommand{\osum}{{%
    \setbox0\hbox{\circ}%
    \rlap{\hbox to \wd0{\hss\sum\hss}}\box0
}}

\renewcommand{\vec}[1]{\mathbf{#1}}
\newcommand{\beq}{\begin{equation}} 

\newcommand{\eeq}[1]{\label{#1} \end{equation}}

\def\LBCO{La$_{2-x}$Ba$_x$CuO$_4$}
\def\YBCO{YBa$_2$Cu$_3$O$_{6+x}$}

\def\BSCCO{Bi$_2$Sr$_2$CaCu$_2$O$_{8+\delta}$}

\def\C60{A$_x$C$_{60}$}

\def\HgCu3{HgCa$_2$Cu$_3$O$_{8+y}$}
\def\HgCu4{HgBa$_2$Ca$_3$Cu$_4$O$_{10+y}$}
\def\TlCu{Tl$_2$Ba$_2$CuO$_{6+\delta}$}
\def\TlCu3{Tl$_2$Ba$_2$Ca$_2$Cu$_3$O$_{10+y}$}
\def\TlCu4{Tl$_2$Ba$_2$Ca$_3$Cu$_4$O$_{12+y}$}

\def\BiCu3{Bi$_2$Sr$_2$Ca$_{2}$Cu$_3$O$_y$}

\def\8LSCO{La$_{1.88}$Sr$_{.12}$CuO$_4$}
\def\110LNSCO{La$_{1.5}$Nd$_{0.4}$Sr$_{0.1}$CuO$_{4}$}
\def\stage4LCO{La$_{2}$CuO$_{4+\delta}$}
\def\Y248{YBa$_2$Cu$_4$O$_8$}

\def\NbSe2{NbSe$_2$}
\def\TaSe2{TaSe$_2$}
\def\TiSe2{TiSe$_2$}
\def\NaCoOH2O{Na$_{0.3}$CoO$_{2y}$H$_2$O}
\def\MgB2{MgB${}_2$}

\def\URu2Si2{URu$_2$Si$_2$}

\def\Ba122{Ba(Fe$_{1-x}$Co$_x$)$_2$As$_2$}

\def\hts{high temperature superconductors}

\begin{document}

\title{Pair-Density-Wave Superconducting States and Electronic Liquid Crystal Phases}   
\author{Rodrigo Soto-Garrido and Eduardo Fradkin}         
\affiliation{Department of Physics and Institute for Condensed Matter Theory, University of Illinois at Urbana-Champaign, 1110 
West Green Street, Urbana, IL  61801-3080, USA} 
\date{\today}    

\begin{abstract}
In conventional superconductors the Cooper pairs have a zero center of mass momentum. In this paper we  present a theory of 
superconducting states where the Cooper pairs have a nonzero center of mass momentum, inhomogeneous superconducting states known as a 
pair-density-waves (PDW) states. 
We show that in a system of spin-1/2 fermions in 2 dimensions in an electronic nematic spin triplet  phase where rotational symmetry is broken 
both in real and in spin space PDW phases arise naturally in a theory that can be analyzed using controlled approximations.
We show that several superfluid phases  that may arise in this phase can be treated within a controlled  BCS mean field theory, with the strength of the spin-triplet
nematic order parameter playing the 
role of the small parameter of this theory.  We find that in a spin-triplet nematic phase, in addition of a triplet $p$-wave and spin-singlet 
$d$-wave (or $s$ depending on the nematic phase) uniform superconducting states, it is also possible to have 
a $d$-wave (or $s$) PDW superconductor. The PDW phases found here can be either unidirectional, bidirectional or tridirectional depending on the
spin-triplet nematic phase and which superconducting channel is dominant. In addition, a triple-helix state is found in a  particular channel.
We show that these PDW phases are present in the weak coupling limit,
in contrast to the usual Fulde-Ferrell-Larkin-Ovchinnikov phases which require strong coupling physics in addition to a large magnetic 
field (and often both). 
\end{abstract}

\maketitle

\section{Introduction}

The problem of the interplay between superconductivity and other broken symmetry states is one of the  central problems in the physics of 
strongly correlated systems.
This issue is particularly pressing in the context of the cuprate {\hts} and their complex phase diagram.  
In addition to N\'eel antiferromagnetic order and   high $T_c$ uniform $d_{x^2-y^2}$ superconductivity, a host of other ordered phases, 
including incommensurate spin stripes (which exhibit spin-density-wave (SDW) order), incommensurate charge stripes 
(with charge-density-wave (CDW) order), electronic nematic order, and time-reversal (and/or mirror-plane) symmetry-breaking  have been 
reported essentially in all the cuprate {\hts}.\cite{kivelson-2003,vojta-2009,Fradkin-2012b,Fradkin-2010}  
Static spin stripe order is seen in  the lanthanum family of the cuprate superconductors.\cite{tranquada-1995,lake-2002,Fujita-2004}
Static charge stripe order is seen in {\LBCO} (LBCO),\cite{abbamonte-2005} in {\YBCO}  (YBCO),\cite{Wu-2011,leboeuf-2012} 
in high magnetic fields
(where otherwise is seen as short range
 order\cite{Ghiringhelli-2012,achkar-2012,chang-2012}), and in {\BSCCO}(BSCCO).\cite{vershinin-2004,howald-2003a,kohsaka-2007}   Nematic charge order is seen in {YBCO}\cite{ando-2002,hinkov-2008,Daou2010} 
and in {BSCCO}\cite{lawler-2010} over a wide range of doping and temperatures. Time-reversal and/or mirror plane (or inversion) symmetry breaking has also 
been reported in {YBCO}, in {LBCO} and in {BSCCO}\cite{Fauque-2006,Xia-2008,kapitulnik-2009,greven-2010} although recent NMR measurements do not detect magnetism in the same samples.\cite{wu-2014}
 Stripe and/or nematic orders of these types are also seen in the iron superconductors\cite{dai-2009,chuang-2010,fisher-2010} and in heavy fermion materials.\cite{matsuda-2011,Park-2012}

A key feature of the orders that are seen in these strongly correlated materials is that the orders are intertwined with each other rather instead of competing with each other.\cite{Berg-2009,Fradkin-2012} 
By intertwined orders what we mean\cite{Fradkin-2010}  is that the orders appear either together and/or with similar strengths, e.g. at critical temperatures of similar magnitude, over a significant range of parameters (doping, coupling constants, etc.) Instead, if the orders were competing 
with each other, one of the orders will be stronger and the others will be strongly suppressed. The exception to this rule are systems which are close to a multicritical point at which not only the critical temperatures but also all the couplings between the different orders are finely-tuned to very specific relations (and values). While this can happen in a particular material at a particular doping it is unnatural to assume that multicriticality should generically occur in all materials and for a wide range of parameters.
 
A case that is particularly relevant from the perspective of intertwined orders is LBCO, particularly near the so-called 1/8 anomaly. In this material the $T_c$ of the uniform $d$-wave superconductivity is suppressed (down to low-temperatures). Yet, a variety of experimental probes show that over essentially the same temperature range where at other dopings LBCO is a $d$-wave superconductors, near 1/8 doping a host of other orders are observed, including charge-stripe order, spin stripe order and a most peculiar phase in which the CuO planes appear to be superconducting but yet the material remains insulating along the c-axis.\cite{Li-2007,tranquada-2008} The layer-decoupling effect is also seen in LBCO away from $x=1/8$ at finite fields\cite{wen-2012} and also in underdoped LSCO materials at finite magnetic fields\cite{schafgans-2010,schafgans-2010b} where a field-induced stripe-ordered state had been observed previously.\cite{lake-2002}

It was suggested by Berg and coworkers that this peculiar layer-decoupling effect 
can be naturally explained if the CuO planes are in an inhomogeneous, striped, superconducting state with the symmetry of 
 a pair-density wave (PDW) state in which charge, spin and superconducting orders are intertwined with each other.\cite{Berg-2007,Berg-2009} 
The local superconducting order parameter  
$\Delta(\bm r)$ in a PDW state is spatially modulated and a spin singlet. For a state with unidirectional modulation, $\Delta(\bm r)$ has the form
\begin{equation}
\Delta(\bm{r})=\Delta_{\bm{Q}}(\bm r)\; e^{i{\bm Q}\cdot\bm{r}}+\Delta_{-\bm{Q}}(\bm r) \; e^{-i{\bm Q}\cdot\bm{r}}
\label{eq:PDW}
\end{equation}
where $\Delta_{\pm {\bm Q}}(\bm r)$ are two slowly-varying complex fields and $\bm Q$ is the ordering wave vector. Hence, the unidirectional PDW superconducting state is characterized by two complex order parameters, $\Delta_{\pm {\bm Q}}(\bm r)$.

A state with the PDW pattern of superconducting order was proposed already in 1964  by Larkin and Ovchinnikov \cite{Larkin-1964} (LO) and by Fulde and Ferrell\cite{Fulde-1964} (FF) to arise in 
the presence of a Zeeman field. As is the case of all ordered phases with a finite wave vector, the LO state, and its time-reversal breaking (spiral)
cousin FF, requires that a nesting condition  be satisfied for this state to occur in the weak 
coupling BCS regime (for a review on FFLO states see Ref.
[\onlinecite{Casalbuoni-2004}]). 
In most cases this nesting conditions is hardly ever satisfied. Thus, states of this type can only exist in a strongly coupled 
regime which is clearly outside the applicability of a weak coupling theory such as BCS. 

Motivated by the LBCO results, and using the BCS framework, Loder and coworkers \cite{Loder-2010} found a PDW state in a tight-binding model with $d$-wave pairing in the absence of an external magnetic field. 
However, these authors found that the  critical value of the coupling constant for which the PDW is the ground state is quite large and hence well outside the regime in which BCS theory is reliable. 
More recently, a PDW state has been found  in variational Monte Carlo simulations of the $t-J$ and $t-t'-J$ model at zero magnetic field,
\cite{Himeda-2002,Raczkowski-2007,Capello-2008,Yang-2008b} although in these simulations  appear to favor the uniform SC state over the PDW state only  by a small amount of 
energy. However, recent, sophisticated iPEPS (infinite projected entangled pair-states\cite{Verstraete-2008}) simulations have found strong evidence for intertwined orders (in which several orders, including the PDW state, appear to be essentially degenerate in energy)  in the $t-J$ model over a significant range of coupling constants and doping.\cite{corboz-2011,corboz-2014} A recent paper by P. A. Lee (which appeared as this work was being finished) suggests that PDW states may arise in a slave-particle RVB approach\cite{lee-2006} by postulating an ``Amperian'' interaction among the spinons.\cite{lee-2014}
On the other hand, a PDW state is known to exist in the spin-gap state of the Kondo-Heisenberg chain\cite{Berg-2010} and also in a two-leg ladder,\cite{Jaefari-2012} even in the weak coupling limit.
 
FFLO states have been proposed to explain some of 
the properties of heavy fermion superconductors \cite{Kenzelmann-2008} and have been conjectured to arise in cold atomic systems. \cite{Radzihovsky-2009} FFLO states were studied in two 
dimensions by Shimahara \cite{Shimahara-1994,Shimahara-1997,Shimahara-1998} where the FFLO states seem to be more robust.

Here we will investigate the relation between PDW states and nematic order. Although charge nematic order (a spatially-uniform spin-singlet state 
that breaks rotational invariance)  does neither favor nor disfavor superconductivity, except in regimes in which $s$-wave and $d$-wave 
superconductivity are in close competition\cite{fernandes-2012,fernandes-2013a,fernandes-2013b} (see, however, Ref.[\onlinecite{kee-2004}]), 
here we will show that a nematic state in the spin triplet channel\cite{Wu-2007} can favor unconventional superconducting phases, including a PDW 
state. In this work we present the study of the presence of an inhomogeneous superconducting  instability in an system that is already in an $\alpha$ or $\beta$ nematic phase. 
We will use a mean field analysis in the weak coupling limit to show that in a region of the phase diagram, an inhomogeneous superconducting state  is the ground state of the 
system.

Oganesyan and coworkers\cite{Oganesyan-2001}  (as well as Refs.[\onlinecite{halboth-2000,khavkine-2004}]) studied a spinless Nematic Fermi fluid 
(FL), where the breaking of rotational symmetry manifest in a spontaneous quadrupolar (elliptical) distortion of the Fermi surface, 
while the translation invariance is preserved (for a review see Ref.[\onlinecite{Fradkin-2010}]). In the charge nematic state the FS has a 
spontaneous quadrupolar (elliptical) distortion. Nematic phases of Fermi fluids can arise either via a Pomeranchuk instability of a Fermi 
liquid\cite{Oganesyan-2001,Fradkin-2010} or by quantum melting of charge stripe phases.\cite{kivelson-1998} 
The resulting anisotropic fluids are non-Fermi liquids if the lattice effects are weak enough.

\begin{figure}[hbt]
\subfigure[]{\includegraphics[width=0.2\textwidth]{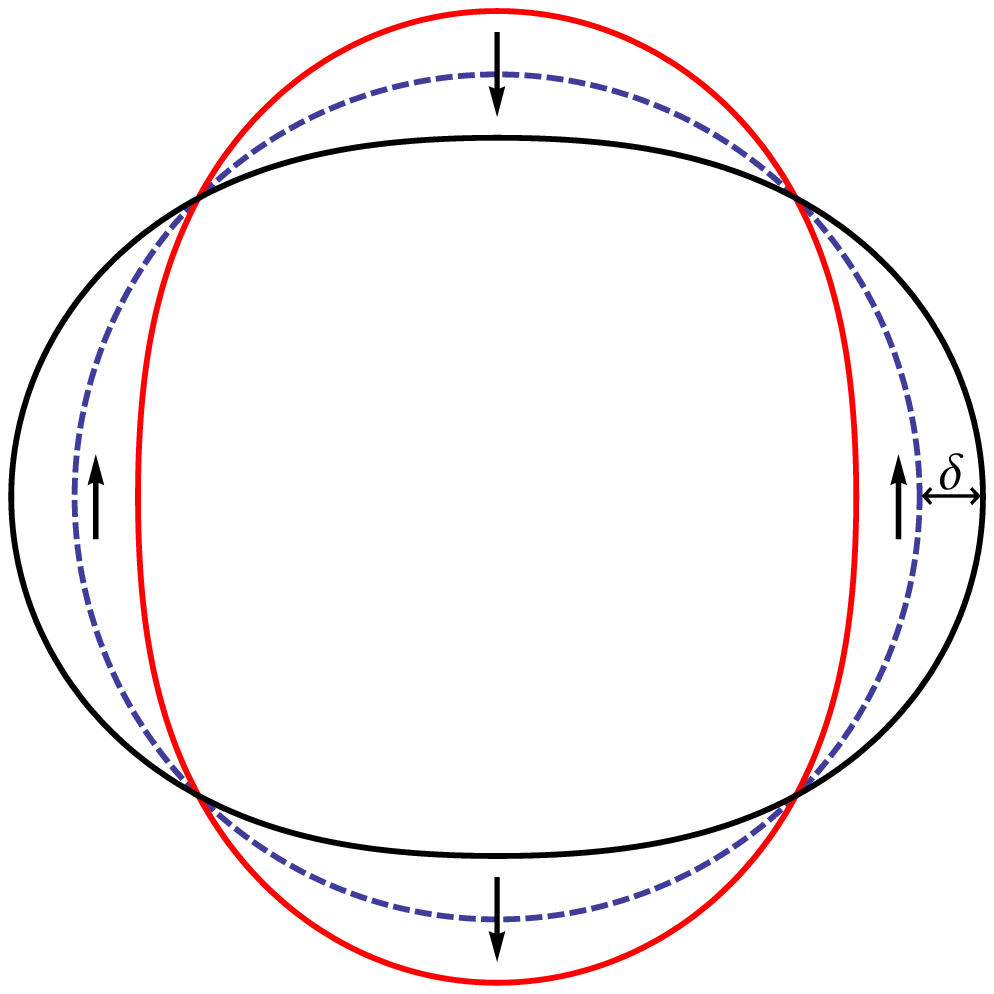}\label{fig:ellipses}}
\subfigure[]{\includegraphics[width=0.2\textwidth]{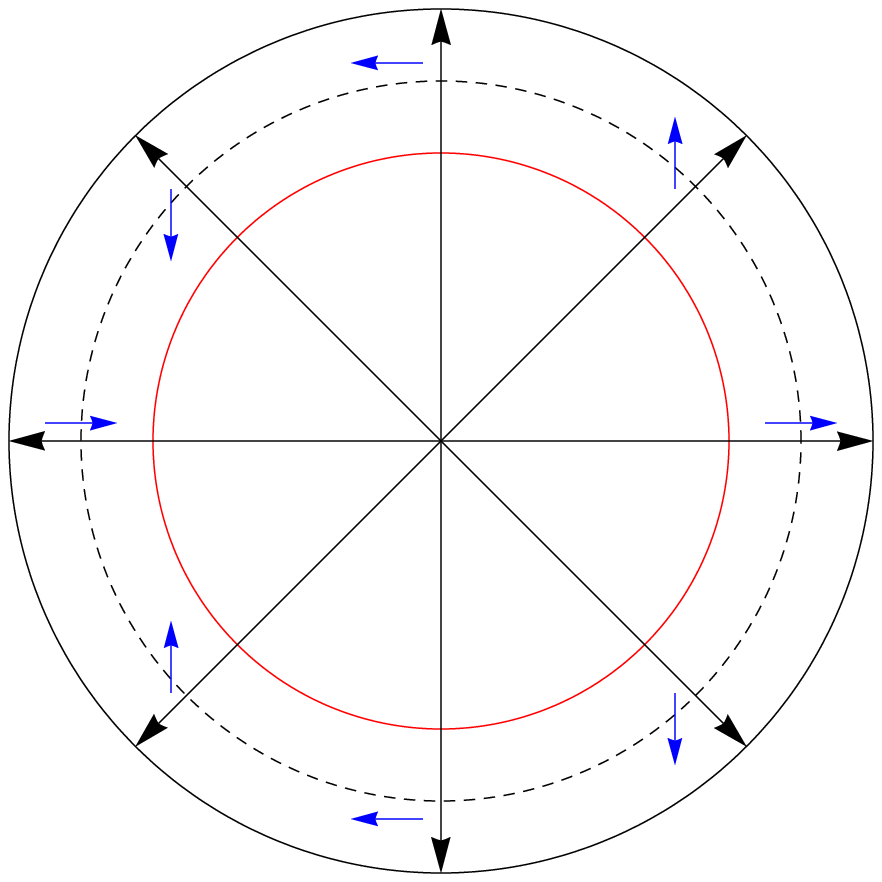}\label{fig:betaw2}}
\subfigure[]{\includegraphics[width=0.2\textwidth]{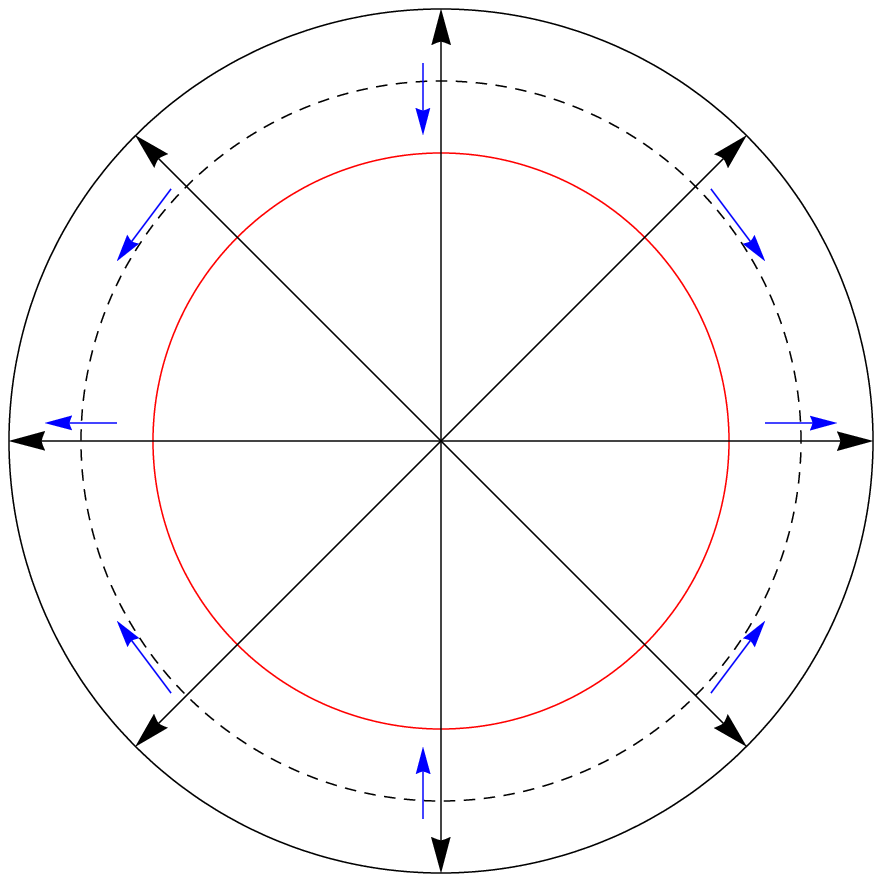}\label{fig:betaw1}}
\subfigure[]{\setlength{\unitlength}{.7cm}\begin{picture}(6,6)
\linethickness{0.3 mm}
\multiput(2,1)(2,0){2}%
{\line(0,2){2}}
\multiput(2,1)(0,2){2}%
{\line(2,0){2}}
\put(1.8,1.7){\vector(1,1.3){.6}}
\put(3.8,1.7){\vector(1,1.3){.6}}
\put(3.4,3.35){\vector(-1,-1.3){.6}}
\put(3.3,1.35){\vector(-1,-1.3){.6}}
\end{picture} \label{YBCO}}
\caption{(color online) (a) Distortion of the FS for the fermions with spin up (black) and down (red) for a triplet $\alpha$ nematic phase with $l = 2$
(Wu et al. \cite{Wu-2007}). 
(b),(c) Distortion of the FS for the fermions with spin plus (black) and minus (red) with respect to the spin quantization axis (blue) for a 
$\beta$ nematic phase with $l = 2$ and $l = -1$ respectively.
(d) Putative nematic spin order in a copper oxide plane in YBCO.\cite{Fauque-2006}}
\end{figure}

Wu et al. \cite{Wu-2007} generalized the aforementioned work of Oganesyan and coworkers to a  system of spin-1/2 fermions and found a 
generalization of the nematic state to the spin triplet channel which they called an $\alpha$-phase. In this phase rotational symmetry is 
broken both in real and in the internal spin space, while while remaining invariant under a combination of a discrete set of rotations in both 
sectors. In addition, they also found another, spatially isotropic  phase, which they called the
$\beta$-phase (in analogy to the $B$ phase in liquid $^3$He). This state is uniform and spatially isotropic, but the spin quantization axis of a 
fermionic quasiparticle on the Fermi surface lies in-plane and winds around the FS with an integer-valued winding number.
In both phases the FS for spin up and down is distorted in different ways (see Figs. \ref{fig:ellipses}, \ref{fig:betaw2}, and \ref{fig:betaw1}) 
providing a natural system to studied the presence of an  instability to an inhomogeneous superconducting state. In a Fermi liquid setting, the 
phase transition to the spin triplet nematic phases occurs as a Pomeranchuk instability and hence the tuning parameter is a Landau parameter in 
the spin triplet channel. In a strong coupling setting it can occur by quantum melting of a spin-stripe state. 
In what follows we will refer to both the $\alpha$ and the $\beta$ phases as spin-triplet nematic phases (although in a strict sense they are not).

In the conventional BCS approach\cite{Fulde-1964,Larkin-1964} the FFLO states arise only in a regime in which there is a sufficiently weak Zeeman 
coupling to an uniform magnetic field so that the SC instability can only occur for Cooper pairs with finite momentum by suppressing the nesting 
between electronic states at the Fermi surfaces for both spin projections. However, this assumption is a severe limitation and, to this date, 
Zeeman-field-tuned  FFLO states have not been clearly seen in experiment. In contrast here we we will see that in the spin triplet nematic phases 
(which although magnetic have a zero uniform Zeeman field) the tuning parameter for the SC instability  is the distance to the nematic spin 
triplet quantum critical point. In particular we will find that depending on whether the nematic is an $\alpha$ or a $\beta$ phase a host of 
different SC states, both uniform and inhomogeneous, can occur.

Unfortunately to this date there is no clear evidence for a spin triplet nematic state. On the theoretical side a recent paper by Maharaj and coworkers\cite{Maharaj-2013} found a spin-triplet $\beta$-phase in a fermionic system on a honeycomb lattice via a Pomeranchuk instability. Fischer and Kim found a nematic-spin-nematic state (the $\alpha$ spin-triplet nematic state) in a mean-field analysis of the three-band Emery model of the cuprates in a regime in which the Hubbard $U_d$ on the Cu sites  and on the O sites ($U_p$) are  both large (and comparable).\cite{fischer-2011}
On the experimental side, there is evidence of time reversal-symmetry-breaking in {\YBCO} close to the
pseudogap temperature in spin-polarized neutron scattering\cite{Fauque-2006,greven-2010} and, with some caveats, 
in Kerr rotation experiments.\cite{Xia-2008,kapitulnik-2009} However, the Kerr rotation experiments can also be interpreted as evidence of inversion symmetry breaking via a gyrotropic effect in a system with charge order.\cite{hosur-2013} Hence the Kerr effect measurements do not on their own prove the existence of as state with broken  time reversal invariance since the cuprate superconductors are now known to exhibit charge order. On the other hand, the spin-polarized neutron experiments can be interpreted either as evidence for loop current order\cite{varma-2005} or as evidence of a nematic spin triplet state which on a CuO lattice means that the oxygens are spin-polarized but their polarization is opposite along the $a$ and $b$ axis (as shown in  Fig. \ref{YBCO}). 
However such a state is incompatible with NMR measurements  which do not find evidence of any sublattice magnetization in YBCO and HBCO which have instead a substantial spin gap.

Aside from these important caveats and reservations, we find that  it is nevertheless useful to consider the possible role of spin triplet nematic phases in a weak-coupling mechanism for pair-density-wave phases. In this work we will consider a system in a spin triplet nematic state but close to the Pomeranchuk quantum critical point. By restricting ourselves to this regime enable us to use controlled approximations. We will assume that the system of interest is inside a  spin-triplet nematic state, sufficiently close to the quantum phase transition so that the magnitude of the order parameter. However we will also assume that we are deep enough in the spin-triplet nematic phase so that the quantum critical fluctuations can be safely ignored. Furthermore we will also ignore the possible non-Fermi liquid physics which may arise in the spin-triplet nematic state.
Thus, the main assumption that we will use throughout is the existence to the Pomeranchuk quantum critical point and that the resulting $\alpha$ and $\beta$ phases are stable. For this reason we will not consider the $\l=1$ case since these phases are unstable in the absence of sufficiently strong spin-orbit interactions.\cite{Wu-2007}
We will show that, depending on the particular spin triplet nematic phase that is considered, different uniform superconducting phases arise ($s$, $p$ or $d$ wave) and  that these phases are in close competition with inhomogeneous phases with the symmetry of a pair-density-wave of the LO type. FF states are generally found to be metastable at least close to the thermal phase boundary. 

The main results of this work are summarized in three phase diagrams, one for the spin triplet nematic $\alpha$ phase with pairing in the 
$d$-wave superconducting channel (shown in Fig.\ref{phasediagram}) and two for the spin triplet nematic $\beta$ phase with pairing in the $s$ 
and $d$ wave superconducting channels (shown in Fig.\ref{fig:beta-diagram-swave} and Fig.\ref{fig:beta-diagram-dwave} respectively.)
We also determine the structure of the Landau-Ginzburg free energies close to the thermal transition and calculate the coefficients and 
stiffnesses. The resulting phase diagrams turn out to be quite complex. In the case of the $\alpha$ phase the superconducting states which arise
are, in addition to a spin-triplet $p$ wave state, a uniform spin singlet $d$-wave SC, a bidirectional PDW state, and a unidirectional PDW state. 
On the other hand, in  the case of the $\beta$ phase the uniform state may be an $s$-wave or a $d$ wave SC. If the pairing channel is $s$ wave, 
in the $\beta$ phase we find unidirectional, bidirectional and tridirectional PDW states and, in addition, a triple-helix FF-type state. If the pairing channel is $d$-wave, in addition to an uniform $d$-wave SC, we also find both a unidirectional and two bidirectional PDW phases.  We also investigate the nature of the phase transitions between these states close to the thermal phase boundary. A rich set of different behaviors are found, including continuous and first order phase transitions as well as Lifshitz points and other multicritical points. It is important to emphasize that these results, obtained using a weak coupling BCS theory, are controlled by the distance to the spin triplet nematic quantum critical point. Thus the spin triplet nematic quantum  critical point plays the role of a complex multicritical point.

This paper is organized as follows. In Section \ref{sec:spin-triplet-nematic} we summarize the theory and description of the spin-triplet nematic phases and follow closely the results and notation of Ref. [\onlinecite{Wu-2007}]. This caveats are discussed in this section in some detail. In Section \ref{sec:SCinst} we discuss the SC instabilities of the $\alpha$ (Subsection \ref{sec:alpha-phase}) and $\beta$ (Subsection \ref{sec:beta-phase}) phases by calculating explicitly the respective SC susceptibilities. In Section \ref{sec:MFT} we present a BCS-type mean-field theory of the different SC states and show that it is well controlled in the regime where the spin-triplet nematic order parameter is small enough. In this Section we derive the Landau-Ginzburg free energy for each phase and derive the phase diagrams and in Section \ref{sec:conclusions} we present our conclusions. The details of the calculations are presented in the Appendix.

\section{Spin-Triplet Nematic Phases}
\label{sec:spin-triplet-nematic}

We start by recalling some of the main results on spin-triplet nematic phases in two dimensions from Ref. [\onlinecite{Wu-2007}] which are relevant for the present work. 
The mean-field (MF) Hamiltonian \cite{Wu-2007} for a spin-triplet nematic phase is:
\begin{align}\nonumber
 H=&\sum_{{\bm k}}c^{\dagger}_{{\bm k},\alpha}\{\epsilon_{{\bm k}}-[{\bm n}_1\cos(l\theta)+
 {\bm n}_2\sin(l\theta)]\cdot {\bm \sigma}_{\alpha,\beta}\}c^{}_{{\bm k},\beta}\\
 &+\frac{|{\bm n}_1|^2+|{\bm n}_2|^2}{2|f_l^a|}
 \label{HMFnematic}
\end{align}
where ${\bm n}_1$ and ${\bm n}_2$ are the order parameters for the spin-triplet nematic phase, ${\bm \sigma}=(\sigma^x,\sigma^y,\sigma^z)$ are the three $2 \times 2$ Pauli matrices, 
$l\in\mathbb{Z}$, $\theta$ is the polar angle between ${\bm k}$ and the $k_x$ axis and $f_l^a$ are the  Landau parameters in the 
spin triplet channel of Fermi liquid theory.\cite{FLBaym} 

The order parameter fields  ${\bm n}_1$ and ${\bm n}_2$ transform under a global $SO(3)_S$ rotation $R$ in the spin channel (denoted here by $S$)    as follows
\begin{equation}
 {\bm n}_1\mapsto R\cdot{\bm n}_1, \qquad {\bm n}_2 \mapsto R\cdot{\bm n}_2
\end{equation}
In addition, the order parameter fields ${\bm n}_1$ and ${\bm n}_2$ transform as follows 
under a spatial rotation by a global angle $\theta$ about the $z$ axis perpendicular to the 2D plane 
\begin{align}
 {\bm n}_1&\mapsto \cos(l\theta){\bm n}_1+\sin(l\theta){\bm n}_2 \nonumber \\
 {\bm n}_2&\mapsto -\sin(l\theta){\bm n}_1+\cos(l\theta){\bm n}_2
 \end{align}
 We will refer to this as the $SO(2)_L$ ``orbital'' (or spatial) rotational invariance. This symmetry is exact in an electron fluid in the 
 continuum and reduces to a discrete subgroup for a lattice model, i.e. the point or space group of the lattice, and it is contained in the 
 symmetries of the free-fermion band structure denoted in Eq.\eqref{HMFnematic} by $\epsilon_{{\bm k}}$. For simplicity in this paper we will 
 consider an electron fluid in the continuum in which case $\epsilon_{\bm k}$ is invariant under $SO(2)_L$ rotations.
 
The Ginzburg-Landau (GL) free energy for the system must be invariant under the global combined  symmetry $SO(2)_L\otimes SO(3)_S$. We will focus first  in the dependence of the GL free energy for phases in which the order parameter fields ${\bm n}_1$ and ${\bm n}_2$ take uniform
values, and hence do not depends on the position ${\bm x}$. Under this assumption, to low orders in the order parameter fields, 
the most general $SO(2)_L\otimes SO(3)_S$-invariant form of the GL free energy is given by:
\begin{align}
 F(&{\bm n}_1,{\bm n}_{2})=\nonumber\\
 =& r(|{\bm n}_1|^2+|{\bm n}_2|^2)+v_1(|{\bm n}_1|^2+|{\bm n}_2|^2)^2+v_2|{\bm n}_1\times{\bm n}_2|^2
 \nonumber\\
 &+\ldots
 \label{GLnematic}
\end{align}
where $r$, $v_1$ and $v_2$ are three parameters (or coupling constants). As usual $r$ is a linear measure of the distance to the critical temperature (for the thermal transition) or to the critical coupling constants (e.g. the Landau parameters $f_l$) in the case of the quantum phase transition.

For $r<0$ the system is in a broken symmetry state, and the  GL free energy in Eq. \eqref{GLnematic} has two type of solutions depending on the 
sign of $v_2$. For $v_2>0$ it is most favorable to have a state where ${\bm n}_1\parallel{\bm n}_2$. This is the  $\alpha$-phase.\cite{Wu-2007} 
On the other hand, for $v_2<0$ it is most favorable to have a state where ${\bm n}_1\perp{\bm n}_2$ and $|{\bm n}_1|=|{\bm n}_2|$. 
This is the  $\beta$-phase.\cite{Wu-2007}

In the  $\alpha$-phase the Fermi surface (FS) of the electrons with spin up and down become spontaneously anisotropic in space.
Hence in this phase both $SO(2)_L$ and $SO(3)_S$ are spontaneously broken symmetries. In this phase, we can choose 
${\bm n}_1=\bar{n}\hat{{\bm z}}$ and ${\bm n}_2=0$  (notice that we can get a non zero ${\bm n}_2$ just doing a rotation around
the $z$ axis, so this is always allowed). 
However, in the $\alpha$ phase the system retains the discrete unbroken symmetry of  spatial rotations by $\pi/l$ combined with a global spin flip. 
On the other hand, the $\beta$-phase corresponds to a phase where the spin polarization axis
winds around the FS. Here we choose $|{\bm n}_1|=|{\bm n}_2|=\bar{n}$ and ${\bm n}_1=\bar{n}\hat{{\bm x}}$ and  ${\bm n}_2=\bar{n}\hat{{\bm y}}$ (which
can always be achieved by a rotation in spin space). 

In the following sections we will  discuss the  SC instabilities (and phases) which arise  in these $\alpha$ and $\beta$ phases.
To this end, in addition to the Hamiltonian in Eq. \eqref{HMFnematic}, we will add a pairing interaction in the spin-singlet channel of the form\cite{Mineev-1999}
\begin{equation}
 H_p=\sum_{{\bm k},{\bm k}',{\bm q}}V({\bm k},{\bm k}')c^{\dagger}_{{\bm k}+{\bm q}/2,\uparrow}c^{\dagger}_{-{\bm k}+{\bm q}/2,\downarrow}
c^{}_{-{\bm k}'+{\bm q}/2,\downarrow}c^{}_{{\bm k}'+{\bm q}/2,\uparrow}
 \label{Interaction}
\end{equation}
where 
\begin{equation}
V({\bm k},{\bm k}')=-g_{\lambda}\gamma_{\lambda}(\hat{{\bm k}})\gamma_{\lambda}(\hat{{\bm k}}')
\end{equation}
where $g_{\lambda}$ is the coupling constant 
in the channel labeled by $\lambda$, and $\gamma_{\lambda}(\hat{{\bm k}})$ is the normalized form factor of the $\lambda$ channel (e.g. $\lambda$ can 
correspond to $s$, $d$, $\dots$ wave pairing) and obey the normalization condition
\begin{equation}
\displaystyle\int \frac{d\theta}{2\pi}\gamma^2_{\lambda}(\hat{{\bm k}})=1
\end{equation}
For instance, the $s$-wave and $d$-wave form factors are
 \begin{align}
 \gamma_{s}(\hat{{\bm k}})=&1,  &(s-\textrm{wave})\nonumber\\
\gamma_{d_{x^2-y^2}}(\hat{{\bm k}})=&\sqrt{2}(\hat{{\bm k}}_x^2-\hat{{\bm k}}_y^2)=\sqrt{2}\cos2\theta &(d-\textrm{wave})
\label{form-factors}
\end{align}
As usual, the $s$-wave form factor is nodeless while the $d$-wave form factor has nodes at $\theta=(2n+1)\pi/4$, where $n \in \mathbb{Z}$. 

We will show below that there are SC instabilities at  critical values of the coupling constants $g_\lambda^c$, which  are controlled (tuned)  
by the expectation value of the spin-triplet nematic order parameter, denoted above by $\bar{n}$ which, in turn, is determined by how far the system is into a spin triplet nematic state from its  quantum critical point to the normal Fermi fluid. In particular we will see that for $\bar n$ 
small enough there are SC instabilities in the weak coupling regimes of these coupling constants. Therefore, the  theory we are presenting in this work can be regarded as a theory of a multicritical 
point for a system close to spin-triplet nematic  phases and superconducting phases (both uniform and non-uniform).

\section{Superconducting instabilities}
\label{sec:SCinst}

We start by looking at the Cooper instability in the $s$-wave and $d$-wave channels for both the $\alpha$- and $\beta$-phases in each of the spin triplet nematic phases. 
We begin by writing down the SC susceptibility (i.e. the bubble diagram in the particle-particle channel) of the isotropic electron fluid 
$ \chi_{sc}({\bm Q},i\omega_m)$, 
\begin{widetext}
\begin{equation}
 \chi_{sc}({\bm Q},i\omega_m)=\displaystyle
T\sum_{n=-\infty}^{\infty}\int \frac{d^2k}{(2\pi)^2} \; \gamma_{\lambda}^2(\hat{{\bm k}})
G_0({\bm k}+{\bm Q}/2,i\omega_n+i\omega_m/2)G_0(-{\bm k}+{\bm Q}/2,-i\omega_n+i\omega_m/2),
\label{susceptibilitymatsu1}
\end{equation}
\end{widetext}
where $\omega_n=(2n+1)\pi T$ are fermionic Matsubara frequencies, 
$\omega_m=2m\pi T$ are bosonic Matsubara frequencies, and 
\begin{equation}
G_0({\bm k},i\omega_n)=\displaystyle\frac{1}{i\omega_n-\epsilon({\bm k})}
\end{equation}
 is the free-fermion Green function.
After performing the Matsubara sum in Eq. \eqref{susceptibilitymatsu1} we obtain
\begin{align}
 \chi_{sc}({\bm Q},i\omega_m)=\quad&\nonumber\\
 =\int \frac{d^2k}{(2\pi)^2}\gamma_{\lambda}^2(\hat{{\bm k}}) & \frac{1-n_F(\epsilon({\bm k}+{\bm Q}/2))-n_F(\epsilon(-{\bm k}+{\bm Q}/2))}
{\epsilon({\bm k}+{\bm Q}/2)+\epsilon({\bm -k}+{\bm Q}/2)-i\omega_m}
\label{susceptibilitymatsu2}
\end{align}
and
\begin{equation}
n_F(\epsilon)=\displaystyle\frac{1}{e^{\epsilon/T}+1}
\end{equation}
 is the Fermi-Dirac distribution.

At finite temperature,  Eq. \eqref{susceptibilitymatsu2} in general has to be evaluated numerically. However, at
 zero temperature it is possible to obtain explicit analytic expressions for the SC susceptibility. Below, we will focus first on the zero temperature
SC instabilities and we will take $\omega_m=0$. In this case we find
\begin{align}
 \chi_{sc}({\bm Q})=\qquad\quad&\nonumber\\
=\int \frac{d^2k}{(2\pi)^2}\gamma_{\lambda}^2(\hat{{\bm k}}) &\frac{1-\Theta(-\epsilon({\bm k}+{\bm Q/2}))-\Theta(-\epsilon(-{\bm k}+{\bm Q/2}))}
{\epsilon({\bm k}+{\bm Q/2})+\epsilon({\bm -k}+{\bm Q/2})}
\label{susceptT0}
\end{align}
We will evaluate Eq. \eqref{susceptT0} for both the $\alpha$- and $\beta$-phases.
\subsection{$\alpha$-phase}
\label{sec:alpha-phase}

From now on we will focus in the (quadrupolar) $l=2$ channel. In this state, the system remains invariant under a spatial rotation of $\pi/2$ 
followed by a global spin flip. The $\alpha$ phase is represented by  the choice ${\bm n}_1=\delta\hat{{\bm z}}$ and ${\bm n}_2=0$. 
Hereafter we  will use the notation $\bar{n}\rightarrow\delta$, 
to explicitly state that in the $\alpha$ phase the Fermi surfaces  of the up and down spin fermions are distorted as shown in 
Fig. \ref{fig:ellipses}), with $\delta$ being the distortion. 
Notice that from Eq. \eqref{GLnematic}, for the $\alpha$-phase ($v_2>0$) we have that:
\begin{align}
 F&=r\delta^2+v_1\delta^4+\ldots
 \label{GLnematicalpha}
\end{align}
which has a minimum at $\delta=\sqrt{|r|/2v_1}$. We can see that $\delta$ scales with the distance to the quantum critical point. Therefore we 
can control $\delta$, controlling the parameter $r$. Keeping that in mind we can write the superconducting susceptibility at wave vector $\bm Q$  in the $\alpha$ 
phase in the SC channel $\lambda$, $\chi_\alpha^\lambda(\bm Q)$,  in the form
\begin{equation}
 \frac{\chi_{\alpha}^\lambda({\bm Q})}{N(E_F)}=
\int_0^{2\pi}\frac{d\theta}{2\pi}\gamma^2_{\lambda}(\hat{{\bm k}})
\ln\left|\frac{\omega_D}{\delta\cos(2\theta)-\frac{Q}{2}\cos(\theta-\phi)}\right|
\label{susc}
\end{equation}
where $\gamma_\lambda({\hat {\bm k}})$ are the form factors for the $s$ and $d$ wave pairing channels defined in Eq.\eqref{form-factors}.
To get the previous expression we have used the notation
\begin{equation}
\int \frac{d^2k}{(2\pi)^2}\rightarrow N(E_F)\int_{-\omega_D}^{\omega_D}d\xi\int_0^{2\pi}\frac{d\theta}{2\pi}
\end{equation}
and then integrated over the excitation energy $\xi$ (measured from the undistorted FS), where $\omega_D$ is an energy cutoff. 
Here $Q$ and $\phi$ are the magnitude and the polar angle of the momentum ${\bm Q}$ (with ${\bm Q}$ being the center of mass momentum of the 
Cooper pairs) and $N(E_F)$ the density of states on the FS, which will be assumed to be constant.

\begin{figure}[t]
 \includegraphics[width=0.48\textwidth]{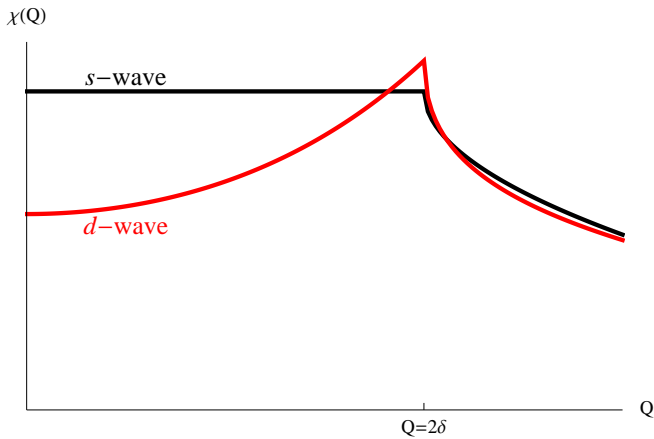}
\caption{(color online) SC susceptibility for the $\alpha$-phase in the $s$-wave and $d$-wave channels in the direction $\phi=n\pi/2$.}
\label{susceptibility}
\end{figure}

The susceptibility for the $s$-wave and $d$-wave channels in the direction $\phi=n\pi/2$ are plotted in Fig. \ref{susceptibility} as a function of $Q=|{\bm Q}|$.
It is important to note here that while in the $s$-wave channel there is no preference for a finite value of $Q$, there is a clear preference in 
the $d$-wave channel, where the susceptibility is enhanced at $Q=2\delta$ and $\phi=n\pi/2$. 
This provides a first evidence that at least at zero temperature an 
inhomogeneous superconducting state  can be the ground state of the system. From now on we will focus on the $d$-wave channel, since we are interested in an inhomogeneous superconducting state. 

The SC susceptibility for $s$ and $d$ channels at ${\bm Q}=0$ are finite in the $\alpha$ spin triplet nematic state and are given by
\begin{align} 
\chi_\alpha^s(0)=&N(E_F) \ln \left(\frac{2\omega_D}{\delta}\right)\nonumber\\
\chi_\alpha^d(0)=&N(E_F) \ln \left(\frac{2\omega_D}{e^{1/2}\delta}\right)
\end{align}
In contrast with the case of a Fermi liquid in which the SC  susceptibilities are divergent at ${\bm Q}=0$ (due to the nesting property of the Fermi surface), in the spin triplet nematic $\alpha$ phase they are finite. Consequently in the $\alpha$ phase there is a (mean field theory) critical value of the pairing coupling constants $g_s$ and $g_d$ for the uniform SC state to occur,
\begin{equation}
g_s^c=\chi_\alpha^s(0)^{-1}, \quad g_d^c=\chi_\alpha^d(0)^{-1}
\end{equation}

The value of the 
susceptibility at $Q=2\delta$ can be determined evaluating Eq. \eqref{susc}, 
\begin{equation}
\chi^d_{\alpha}({\bm Q}_{op}) =N(E_F) \ln\left(\frac{2\omega_D e^{1/8}}{\delta}\right)
\end{equation}
 where $|{\bm Q}_{op}|=2\delta$ and ${\bm Q}_{op}$ 
points in the $n\pi/2$ direction. 
The (mean field theory) critical value of the coupling constant in order to have a Cooper instability at finite ${\bm Q}$ in the $d$ wave channel is 
\begin{equation}
 g^{\alpha}_{d_c}({\bm Q}_{op})=\chi^d_\alpha({\bm Q}_{op})^{-1}
\label{dwaveQcriticalcoupling}
\end{equation}
In the $d$-wave case there is an extra factor of $e^{1/8}$ that is not present in the $s$-wave channel. 
This extra factor reduces the critical value of the coupling constant in the $d$-wave channel. 

An important feature of the result of Eq.\eqref{dwaveQcriticalcoupling} is that the value of $g_{d_c}$ is controlled by the magnitude 
${\bar n}=\delta$ of the spin-triplet nematic state which, more geometrically, parametrizes the distortions  $\delta$ of the Fermi surfaces for 
fermions with up and down spins. 
It is the smallness of the parameter $\delta$ that allows us to work in the weak coupling regime and hence to use BCS theory when $\delta$ is 
very small. 
This result will be extended in the next Section to finite temperature where it will be used to determine the phase diagram.

Finally let us discuss briefly the role of  spin-triplet pairing interactions
(e.g. $p$-wave pairing). In contrast to what we found in the singlet $s$ and $d$ wave channels,  the Fermi surfaces of the $\alpha$ phase are still nested. As a result, there is an infinitesimal SC instability in the uniform $p$-wave channel.
 However, provided we  assume that the coupling constant for this pairing channel is sufficiently weak,  the $T_c$ for the $d$-wave 
channel is always higher than the $T_c$ for the $p$-wave channel. In what follows we will ignore the $p$-wave channel.
 
In conclusion, in the $\alpha$ phase there is a critical value of the pairing coupling constant for both  the $s$- and $d$-wave uniform SC channels. However, the $s$-wave channel does not favor the formation of SC states with finite wave vector  whereas the $d$-wave channel clearly does, as shown in Fig.\ref{susceptibility}. In what follows we will only consider the case of the $d$-wave channel.

\subsection{$\beta$-Phase}
\label{sec:beta-phase}
From Eq. \eqref{GLnematic} for the $\beta$-phase ($v_2<0$) we have that:
\begin{align}
 F&=2r\bar{n}^2+4v_1\bar{n}^4+v_2\bar{n}^4+\ldots
 \label{GLnematicabeta}
\end{align}
which has a minimum at $\bar{n}=\sqrt{|r|/{(4v_1+v_2)}}$. We can see that $\bar{n}$ scales with the distance to the quantum critical point. 
Therefore, we can control $\bar{n}$, controlling the parameter $r$. 
As for the $\alpha$-phase we start by looking at the Cooper instability in the $\beta$-phase.   Since in the $\beta$-phase this case the FS's 
are spherically symmetric 
(see Figs. \ref{fig:betaw2} and \ref{fig:betaw1}), the SC susceptibility in the pairing channel $\lambda$ at finite temperature $T$ (Eq. \eqref{susceptibilitymatsu2} with 
$\omega_m=0$) can be written for general $l$ as: 
\begin{widetext}
\begin{align}\nonumber
 \frac{\chi_{\beta}^\lambda({\bm Q},T)}{N(E_F)}=&\int_{-\omega_D}^{\omega_D}d\xi\int_0^{2\pi}\frac{d\theta}{2\pi}\gamma^2_{\lambda}(\hat{{\bm k}})
 \frac{1}{8\xi(\bar{n}-\xi )(\bar{n}+\xi )}\\ \nonumber
 &\left[\left(\bar{n}(-1)^l+\bar{n}-2\xi \right)(\bar{n}+\xi )\left( \tanh \left(\frac{\bar{n}-\xi -Q/2 \cos(\theta-\phi)}{2 T}\right)+
 \tanh \left(\frac{\bar{n}-\xi +Q/2 \cos(\theta-\phi)}{2T}\right)\right)\right.\\
 &-\left.\left(\bar{n}(-1)^l+\bar{n}+2\xi \right)(\bar{n}-\xi )\left( \tanh \left(\frac{\bar{n}+\xi -Q/2 \cos(\theta-\phi)}{2 T}\right)+
 \tanh \left(\frac{\bar{n}+\xi +Q/2 \cos(\theta-\phi)}{2T}\right)\right)\right]
 \label{suscbeta}
\end{align} 
\end{widetext}
Notice that the expression for the the SC susceptibility in Eq. \eqref{suscbeta} depends only on the parity of $l$, and not
on it's value. 

Let us analyze briefly the behavior of the SC susceptibilities for the $l$ odd and $l$ even cases before discussing  the zero temperature limit.

\subsubsection{l odd}

For $l$ odd the expression of the SC susceptibility in pairing channel $\lambda$ of Eq. \eqref{suscbeta} reduces to:
\begin{widetext}
\begin{align}\notag
\frac{\chi_{\beta}^\lambda({\bm Q},T)}{N(E_F)}=\frac{1}{4}\int_{-\omega_D}^{\omega_D}d\xi\int_0^{2\pi}\frac{d\theta}{2\pi}\gamma^2_{\lambda}(\hat{{\bm k}})
 &\left[\frac{1}{(\bar{n}-\xi )}\left(\tanh \left(\frac{\bar{n}-\xi -Q/2 \cos(\theta-\phi)}{2 T}\right)+\tanh
   \left(\frac{\bar{n}-\xi +Q/2 \cos(\theta-\phi)}{2 T}\right)\right)\right.\\
   &\left.+   \frac{1}{(\bar{n}+\xi )}\left( \tanh   \left(\frac{\bar{n}+\xi -Q/2 \cos(\theta-\phi)}{2 T}\right)+ 
   \tanh\left(\frac{\bar{n}+\xi +Q/2 \cos(\theta-\phi)}{2 T}\right)\right)\right]
 \label{suscbetalodd}
\end{align} 
\end{widetext}
At $Q=0$ the previous expression reduces to the BCS result
\begin{equation}
 \frac{\chi_{\beta}^\lambda(0,T)}{N(E_F)}
 =\int_{-\omega_D}^{\omega_D} d\xi \; \frac{1}{\xi}\tanh \left(\frac{\xi}{2 T}\right)
 \label{suscbetaloddQ0}
\end{equation}
where we used that $\omega_D\gg\xi$ and we made a change of variables. 
We can then deduce  that for odd $l$, the uniform SC state is the most favorable state
since there is a logarithmic divergence of its susceptibility at $T=0$. Notice the close similarity, for example, 
with the case where there is a finite spin-orbit coupling (see Ref.[\onlinecite{Agterberg}] and references therein). In the case of a Rashba spin-orbit 
interaction (which is similar to the $\beta$-phase with $l=1$), the uniform SC state is favorable in the absence of magnetic field. However, 
in the presence of a Zeeman coupling to a magnetic field, it is possible to favor an inhomogeneous superconducting  state (we will not study the effect of magnetic
fields in the present paper). Those states have been recently studied extensively by Zhang et. al. 
\cite{ZhangPRL-2013,ZhangPRA-2013,ZhangNature-2013}

\subsubsection{l even}

In this case, the SC susceptibility of Eq. \eqref{suscbeta} reduces to:
\begin{widetext}
\begin{align}
 \frac{\chi_{\beta}^\lambda({\bm Q},T)}{N(E_F)}
   =\int_{-\omega_D}^{\omega_D}d\xi\int_0^{2\pi}\frac{d\theta}{2\pi}\gamma^2_{\lambda}(\hat{{\bm k}})
   \frac{1}{4\xi}& \Big[1-\bar{n}_F(\xi+n-Q/2 \cos(\theta-\phi))-n_F(\xi-\bar{n}+Q/2 \cos(\theta-\phi)) \nonumber\\
   & +1-n_F(\xi+\bar{n}+Q/2 \cos(\theta-\phi))-n_F(\xi-\bar{n}-Q/2 \cos(\theta-\phi))\Big]
 \label{suscbetaleven}
 \end{align}
\end{widetext}
Having determined the expression for finite $T$, we will compute the SC susceptibility at $T=0$. After integrating over $\xi$ in Eq. 
\eqref{suscbetaleven} and taking the $T \to 0$ limit, we get:
\begin{equation}
 \frac{\chi_{\beta}^\lambda({\bm Q},0)}{N(E_F)}=\int_0^{2\pi}\frac{d\theta}{2\pi}\gamma^2_{\lambda}(\hat{{\bm k}})
 \ln\left|\frac{\omega_D}{\bar{n}-Q/2 \cos(\theta-\phi)}\right| 
 \label{suscgeneralevenT0}
\end{equation}
For the $s$-wave case, $\gamma_{s}(\hat{{\bm k}})=1$, and  the previous expression can be easily evaluated to be
\begin{align}
 \frac{\chi_\beta^s({\bm Q},0)}{N(E_F)}&=
  \begin{cases}
\displaystyle\ln\left(\frac{2\omega_D/\bar{n}}{1+\sqrt{1-\left({Q/2\bar{n}}\right)^2}}\right), & 0\leq Q\leq 2\bar{n}\\ 
\displaystyle \ln \left(\frac{4\omega_D}{Q}\right), & Q> 2\bar{n}
\end{cases}
  \label{suscbetalevenT0}
\end{align} 
\begin{figure}[t]
 \includegraphics[width=0.48\textwidth]{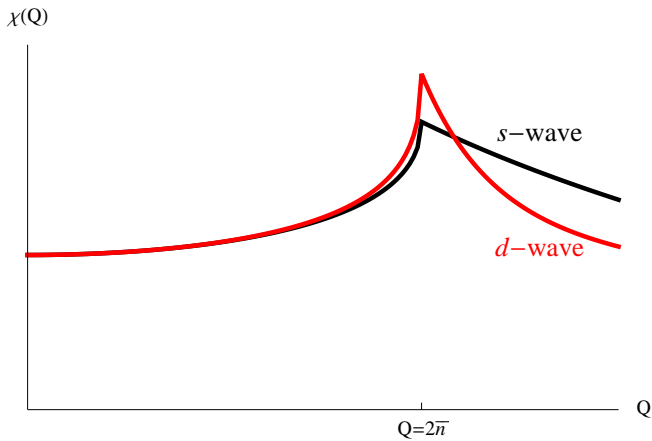}
\caption{(color online) SC susceptibility for the $\beta$-phase in the $s$-wave and $d$-wave channels in the direction $\phi=n\pi/2$.}
\label{susceptibilitybeta}
\end{figure}
We can see that the value of $Q$ that gives the maximum susceptibility is $Q=2\bar{n}$ and ${\bm Q}$ can point in any direction by
rotational symmetry. 

On the other hand, for the $d$-wave case, whose form factor is $\gamma_{d}(\hat{{\bm k}})=\sqrt{2}\cos2\theta$, we have to computed numerically the SC susceptibility of Eq.\eqref{suscgeneralevenT0}, and found that the maximum is at $\phi=n\pi/2$ and $Q=2\bar{n}$, i.e. the antinodal directions
of the $d$-wave order parameter. 

Just as in the case of the $\alpha$-phase, there is a critical value for the pairing coupling constant 
in the $s$- and $d$-wave channels given by the inverse of the respective SC susceptibilities (e.g. Eq.\eqref{suscgeneralevenT0}).
Even though there is a critical value for the coupling constants, this is smaller than the critical value for $Q=0$. 
Therefore, for even $l$, the condensation Cooper pairs with finite momentum  is more favorable (at least at low temperatures for both 
the $s$- and $d$-wave channels). 
Also notice that, as in the $\alpha$-phase, in the $\beta$-phase we also find that  the critical pairing coupling constants in the $s$ and $d$ wave channels obey 
$g_{\beta,c}^{d}<g_{\beta,c}^{s}$ since the $d$-wave channel has a larger SC susceptibility than the $s$ wave channel at the ordering wave vector.
Let us mention that basically the same expression for the susceptibility for the $s$-wave Eq. \eqref{suscbetalevenT0} was obtained by 
Shimahara \cite{Shimahara-1994} who considered an FF phase in a 2D electron gas in the presence of a Zeeman coupling to a perpendicular magnetic field, $h$. 
His expression for the susceptibility differs from us in that our $\bar{n}$ is replaced in his expression by $h$. 
At the mean-field level there is a close analogy between the two problems. 
Here, we can get an inhomogeneous superconducting  phase without an external magnetic field, if we have the system in a $\beta$-phase with even angular momentum $l$.

\section{Mean Field Theory at $T>0$}
\label{sec:MFT}

We will now consider the mean-field (MF) theory of a Hamiltonian that includes the nematic phase and the pairing interaction 
Eq.\eqref{Interaction}. 
For that, we will work in the imaginary time path integral formalism where the action is given by\cite{FradkinFieldTheory} 
\begin{equation}
S=\int_0^{\beta}d\tau \left[ \int d{\bm x}\; \bar{\psi}_\sigma({\bm x}, \tau) (\partial_{\tau}-\mu)\psi_\sigma({\bm x}, \tau)+ H(\bar{\psi},\psi) \right]
\label{eq:action}
\end{equation}
where $\psi_\sigma({\bm x}, \tau)$ is a Fermi field for spin-1/2 fermions, $\mu$ is the chemical potential, and $H$ is the full Hamiltonian. 
We will perform a Hubbard-Stratonovich transformation to get rid of the quartic fermionic terms in the pairing term in $H$. We will consider both the $\alpha$-phase and the $\beta$-phase of the spin-triplet nematic state.

\subsection{$\alpha$-phase}
\label{sec:alpha-MF}

Let us start by
looking at the $\alpha$-phase. In this case the effective action for the superconducting state is given by:
\begin{align}
\begin{split}
  S=&\int_0^{\beta}d\tau\left[\sum_{{\bm k},\sigma}\bar{\psi}_{{\bm k},\sigma}(\partial_{\tau}+\xi_{{\bm k},\sigma})\psi_{{\bm k},\sigma} 
+\sum_{{\bm q}}\frac{|\Delta_{{\bm q}}|^2}{g}\right.\\
&\left.\qquad\qquad-\sum_{{\bm q}}\sum_{{\bm k}}\gamma(\hat{{\bm k}})\bar{\psi}_{{\bm k}+{\bm q}/2,\uparrow}\bar{\psi}_{-{\bm k}+{\bm q}/2,\downarrow}
\Delta_{{\bm q}}\right.\\ 
&\left. \qquad\qquad -\sum_{{\bm q}}\sum_{{\bm k}}\gamma(\hat{{\bm k}})\Delta^*_{{\bm q}}\psi_{-{\bm k}+{\bm q}/2,\downarrow}\psi_{{\bm k}+{\bm q}/2,
\uparrow} \right]
 \label{action}
\end{split}
\end{align}
where $\Delta_{{\bm q}}(\tau)$ is the Hubbard-Stratonovich field associated with the superconducting order parameter at wave vector ${\bm q}$. In the $\alpha$-phase the kinetic energies of fermions with up and down spins measured from  their respective Fermi surfaces are  
\begin{equation}
\xi_{{\bm k},\uparrow}=\xi-\delta\cos2\theta, \quad 
\xi_{{\bm k},\downarrow}=\xi+\delta\cos2\theta
\end{equation}
 respectively, where we have included the magnitude of the spin-triplet nematic order parameter  $\delta$ in the definition of 
 $\xi_{\uparrow,\downarrow}$,
and $\xi$ is the energy measured from the undistorted circular FS.

As we saw
in the Section \ref{sec:SCinst}, there are four equivalent directions for which the SC susceptibility for the $\alpha$-phase has a maximum, 
it is natural to focus in the following four different cases for the superconducting order parameters: 
Fulde-Ferrell (FF), PDW (or Larkin-Ovchinnikov (LO)), bidirectional PDW (or ``checkerboard''), and uniform:

\begin{itemize}
\item
Uniform phases: In the regime in which the $\alpha$-phase order parameter is very small we  find  conventional $p_x$ (or $p_y$) wave (spin-triplet) or $d_{x^2-y^2}$-wave (spin singlet) (depending on which coupling constant is stronger). 

 \item 
 FF phase: In this phase only one wave vector contributes to the SC order parameter 
\begin{equation}
 \Delta({\bm r})=\Delta_{{\bm Q}}(\bm r) e^{i{\bm Q}\cdot{\bm r}}
\label{FF}
\end{equation}
In this phase translation and gauge invariance as well as time reversal and parity are spontaneously broken. 
The SC order parameter field is a one-component complex field $\Delta_{\bm Q}(\bm r)$ (which has a constant expectation value).

 \item 
 PDW phase: two wave vectors contribute to the SC order parameter
\begin{align}
\Delta({\bm r})&=\Delta_{{\bm Q}}(\bm r) \; e^{i{\bm Q}\cdot{\bm r}}+\Delta_{-{\bm Q}}(\bm r)\; e^{-i{\bm Q}\cdot{\bm r}}
\label{PDW}
\end{align}
This state breaks translation and gauge invariance but it is time-reversal invariant. The order parameter field now has two complex components, $\Delta_{\pm {\bm Q}}(\bm r)$ and, hence, has two amplitude fields  $|\Delta_{\pm {\bm Q}}(\bm r)|$ 
and two phase fields, $\theta_{\pm {\bm Q}}(\bm r)=\textrm{arg} [\Delta_{\pm {\bm Q}}(\bm r)]$. In the London gauge and with a choice of origin, and with 
parity invariance $\Delta_{{\bm Q}}=\Delta_{-{\bm Q}}$, the expectation value of the order parameter takes the LO  sinusoidal dependence on 
position, i.e. $\Delta(\bm r)=2|\Delta_{{\bm Q}}|\; \cos({\bm Q}\cdot{\bm r})$. The thermal fluctuations of the phase fields 
$\theta_{\pm {\bm Q}}$ play a key role of the thermal melting of the PDW phase.\cite{Berg-2009b}
 \item 
 Bidirectional phase (or checkerboard) (Bi): in this phase four wave vectors contribute to the SC order parameter,
\begin{equation}
\Delta({\bm r})=\Delta_{{\bm Q}}e^{i{\bm Q}\cdot{\bm r}}+\Delta_{-{\bm Q}}e^{-i{\bm Q}\cdot{\bm r}}
+\Delta_{\bar{{\bm Q}}}e^{i\bar{{\bm Q}}\cdot{\bm r}}+\Delta_{-\bar{{\bm Q}}}e^{-i\bar{{\bm Q}}\cdot{\bm r}}
\label{bidirectional}
\end{equation}
In this phase the SC order parameter is then a four-component complex field with $\Delta_{\pm {\bm Q}}(\bm r)$ and 
$\Delta_{\pm \bar {\bm Q}}(\bm r)$ being the four complex components (and hence four amplitudes and four phase fields). 
Under the assumption of parity and $C_{4}$ symmetry it reduces to
\begin{equation}
\Delta(\bm r)=2|\Delta_{{\bm Q}}| \; (\cos({\bm Q}\cdot{\bm r})+\cos(\bar{{\bm Q}}\cdot{\bm r}))
\end{equation}
where ${\bm Q}\cdot{\bm \bar{\bm Q}}=0$ and we have assumed $|\Delta_{{\bm Q}}|=|\Delta_{-{\bm Q}}|=|\Delta_{{\bm \bar{\bm Q}}}|=|\Delta_{-{\bm \bar{\bm Q}}}|$. 
\end{itemize}
In addition to the four possible states aforementioned, it is also possible to have 2 more states that satisfy the symmetries of the
problem (although as we will show below, and as the FF state, they do not appear in the phase diagram):
\begin{itemize}
 \item 
 Double-helix (2H): in this phase two wave vectors contribute to the SC order parameter,
\begin{equation}
\Delta({\bm r})=\Delta_{{\bm Q}}e^{i{\bm Q}\cdot{\bm r}}+\Delta_{\bar{{\bm Q}}}e^{i\bar{{\bm Q}}\cdot{\bm r}}
\label{doublehelix}
\end{equation}
As in the FF state, in this phase translation and gauge invariance as well as time reversal and parity are spontaneously broken. 
\item 
Bidirectional time-reversal breaking PDW (Bi2): in this phase four wave vectors contribute to the SC order parameter,
\begin{equation}
\Delta({\bm r})=\Delta_{{\bm Q}}e^{i{\bm Q}\cdot{\bm r}}+\Delta_{-{\bm Q}}e^{-i{\bm Q}\cdot{\bm r}}
+\Delta_{\bar{{\bm Q}}}e^{i\bar{{\bm Q}}\cdot{\bm r}}+\Delta_{-\bar{{\bm Q}}}e^{-i\bar{{\bm Q}}\cdot{\bm r}}
\end{equation}
In this phase the SC order parameter is then a four-component complex field with $\Delta_{\pm {\bm Q}}(\bm r)$ and 
$\Delta_{\pm \bar {\bm Q}}(\bm r)$ being the four complex components (and hence four amplitudes and four phase fields). 
In contrast to the bidirectional phase we can take a different choice for the relative phases of the order parameters 
(this corresponds to the phase `5' discussed in Ref. [\onlinecite{Agterberg-2008}]).
\begin{equation}
\Delta(\bm r)=2|\Delta_{{\bm Q}}| \; (\cos({\bm Q}\cdot{\bm r})+i\cos(\bar{{\bm Q}}\cdot{\bm r}))
\end{equation}
This phase breaks time-reversal invariance.
\end{itemize}

Below we will compute the free energy for each one of these phases.

The free energies of the different states are obtained by integrating out the fermionic degrees of freedom in Eq. \eqref{action}. 
For the case of FF phase (and for the uniform phases) it is possible to get an explicit expression for the effective free energy as function of 
the (constant) value of the order parameter field.
However, for the PDW, the bidirectional PDW, the double-helix and time-reversal breaking bidirectional PDW this has to be done numerically 
except near the phase boundary where, if the transition is continuous, the Landau-Ginzburg free energy can be calculated as usual as an 
expansion in powers of the order parameters.

After writing the fermion operators in the Nambu spinor representation
\begin{equation}
\bar{\Psi}_{{\bm k}}=(\bar{\psi}_{{\bm k}+{\bm Q}/2,\uparrow},\psi_{-{\bm k}+{\bm Q}/2,\downarrow})
\end{equation}
 the  action for the general state with a static order parameter $\Delta_{\bm Q}$ has the form
\begin{align}
S_{\rm eff}[\Delta_{\bm Q}]&=-
\sum_{{\bm k},{\bm k}',n} 
{\bar \Psi}_{{\bm k},n} \mathcal{G}^{-1}_{{\bm k},{\bm k}',n} \Psi_{{\bm k}',n}+\beta  \sum_{\bm Q}\frac{|\Delta_{\bm Q}|^2}{g}\nonumber\\
&+\beta\sum_{{\bm k}}\xi_{-{\bm k}+{\bm Q}/2,\downarrow}
\end{align}
In the case of the  FF phase the modes $\Psi_{{\bf k},n}$ with  wave vector ${\bm k}$ and Matsubara frequency $\omega_n$ decouple from each other 
and as a result the matrix $\mathcal{G}^{-1}_{{\bm k},{\bm k}',n}$ is block diagonal. However, this is not the case for the other 
inhomogeneous SC phases aforementioned in which, due to this mixing, it is not possible to write  the free energy in closed form. 
Nevertheless sufficiently close to the phase boundary with the normal state, 
the free energy of the $\alpha$ phase for the inhomogeneous SC phases can be computed perturbatively in powers of the SC order parameter with each 
term being represented by a Feynman diagram computed in the normal phase. Here we will focus only on the phases which arise very close to the 
thermodynamic transition from the normal state. Other phases may occur far from this phase boundary and will not be considered here.

\subsubsection{Free energy of the FF phase}

In the FF case, we can write the action in Eq. \eqref{action} in the simpler form
\begin{align}
 S[\bar{\Psi},\Psi,\Delta_{{\bm Q}},\Delta^*_{{\bm Q}}]=&-
\sum_{{\bm k},n} \bar{\Psi}_{{\bm k},n}\mathcal{G}^{-1}_{{\bm k},i\omega_n}\Psi_{{\bm k},n}+\beta\frac{|\Delta_{{\bm Q}}|^2}{g}\nonumber \\
&+\beta\sum_{{\bm k}}\xi_{-{\bm k}+{\bm Q}/2,\downarrow},
\end{align}
where $\beta=1/T$. Here we  assumed that $\Delta_{\bm Q}$ is constant and real, and  we have used the notation
\begin{equation}
\mathcal{G}^{-1}_{{\bm k},i\omega_n}=
\displaystyle\left( \begin{array}{cc}
i\omega_n-\xi_{{\bm k}+{\bm Q}/2,\uparrow} & \Delta_{{\bm Q}}\gamma(\hat{{\bm k}}) \\
\Delta^*_{{\bm Q}}\gamma(\hat{{\bm k}}) & i\omega_n+\xi_{-{\bm k}+{\bm Q}/2,\downarrow} \end{array} \right)
\label{Green}
\end{equation}
for the inverse of the fermion Green function in the FF phase, where $\gamma(\hat {\bf k})$ is the form factor for the different SC channels. 

After integrating-out the fermionic degrees of freedom we get
\begin{equation}
 S_{\text{eff}}[\Delta_{{\bm Q}},\Delta^*_{{\bm Q}}]=-\ln \det[\mathcal{G}^{-1}]+\beta\frac{|\Delta_{{\bm Q}}|^2}{g}
+\text{const.},
\label{effectiveaction}
\end{equation}
We need to compute
\begin{equation}
\ln \det[\mathcal{G}^{-1}]=\sum_{{\bm k},n} \ln (\lambda^{(1)}_{{\bm k},n}\lambda^{(2)}_{{\bm k},n})
\end{equation}
 where $\lambda^{(i)}_{{\bm k},n}$ are the eigenvalues of the matrix
$\mathcal{G}^{-1}_{{\bm k},i\omega_n}$.
Using that 
\begin{align}
\xi_{{\bm k}+\frac{\bm Q}{2},\uparrow}=&\xi-\delta\cos2\theta+\frac{Q}{2}\cos(\theta-\phi)\nonumber\\
\xi_{-{\bm k}+\frac{\bm Q}{2},\downarrow}=&\xi+\delta\cos2\theta-\frac{Q}{2}\cos(\theta-\phi)
\end{align}
 we find that:
\begin{align}
\lambda^{(1,2)}_{{\bm k},n}=&\pm E_{{\bm k}}+\frac{Q}{2}\cos(\theta-\phi)-\delta\cos2\theta-i\omega_n
\label{evs}
\end{align}
where $E_{{\bm k}}=\sqrt{\xi^2+\gamma^2(\hat{{\bm k}})|\Delta_{{\bm Q}}|^2}$

After integrating-out the fermionic fields we find 
 \begin{widetext}
\begin{align}
\begin{split}
 F_s-F_n=\frac{|\Delta_{{\bm Q}}|^2}{g}-2TN(E_F)\int_{0}^{2\pi}\frac{d\theta}{2\pi}\int_{0}^{\omega_D}d\xi &\left[\ln\left(
\frac{1+e^{-(\sqrt{\xi^2+2\cos^2 2\theta|\Delta_{{\bm Q}}|^2}+Q/2\cos(\theta-\phi)-\delta\cos2\theta)/T}}
{1+e^{-(\xi+Q/2\cos(\theta-\phi)-\delta\cos2\theta)/T}}\right)\right.\\
&\left. +\ln\left(\frac{1+e^{(\sqrt{\xi^2+2\cos^2 2\theta|\Delta_{{\bm Q}}|^2}-Q/2\cos(\theta-\phi)+\delta\cos2\theta)/T}}
{1+e^{(\xi-Q/2\cos(\theta-\phi)+\delta\cos2\theta)/T}} \right) \right]
\end{split}
\label{freenergySC}
\end{align}
\end{widetext}

We can now look for the minimum of the Free energy $F_s$ of Eq.\eqref{freenergySC} with respect to $\Delta_{\bm Q}$ and ${\bm Q}$ to find 
the thermodynamically stable state. We do this minimization numerically over a range of values for $T$ and $\delta$. 
For the $d$-wave channel we find a range of $T$ and $\delta$ in which there is superconducting order, $\Delta\neq0$, but which may also be 
inhomogeneous and hence has  ${\bm Q}\neq0$, as  expected from the SC instabilities computed in section 
\ref{sec:SCinst}. This result suggests the possible presence of either a time-reversal breaking inhomogeneous SC state or a time-reversal 
invariant PDW (or LO) SC state. In addition, the transition from the normal (non-SC state) to the putative FF state 
is continuous. Since a continuous transition is reflected in the divergence of the susceptibility and this is independent of the nature of 
the inhomogeneous SC state, as it is the same for FF, double-helix, unidirectional PDW, bidirectional PDW and time-reversal breaking 
bidirectional PDW, we need to investigate which one of these states actually has lower free energy. 
Since the phase transition is continuous we can investigate the stability of the different phase by  expanding the free energy in powers of 
$\Delta_{\bm Q}$ up to fourth order.

\subsection{Ginzburg Landau Free Energy}
\label{sec:GL-free-energy}

Considering all the possible SC aforementioned phases the most general expression for the free energy compatible with gauge invariance, 
translation invariance and rotation invariance (or point group symmetry) has the form
\begin{align}
 F&=\frac{c_2}{2}\left(|\Delta_{{\bm Q}}|^2+|\Delta_{-{\bm Q}}|^2+|\Delta_{\bar{{\bm Q}}}|^2+
 |\Delta_{-{\bar{\bm Q}}}|^2 \right)\nonumber\\
 &+\frac{c_4}{4}\left(|\Delta_{{\bm Q}}|^4+|\Delta_{-{\bm Q}}|^4+|\Delta_{\bar{{\bm Q}}}|^4+
 |\Delta_{-{\bar{\bm Q}}}|^4 \right)\nonumber\\
 &+\frac{u}{4}\left(|\Delta_{{\bm Q}}|^2|\Delta_{-{\bm Q}}|^2+|\Delta_{\bar{{\bm Q}}}|^2|\Delta_{-{\bar{\bm Q}}}|^2 \right)\nonumber\\
 &+\frac{v_1}{4}\left(|\Delta_{{\bm Q}}|^2|\Delta_{\bar{{\bm Q}}}|^2+|\Delta_{-{\bm Q}}|^2|\Delta_{-{\bar{\bm Q}}}|^2 \right.\nonumber\\
 &\left.\qquad\quad +|\Delta_{{\bm Q}}|^2|\Delta_{-\bar{{\bm Q}}}|^2+|\Delta_{-{\bm Q}}|^2|\Delta_{{\bar{\bm Q}}}|^2\right)\nonumber\\
 &+\frac{v_2}{4}\left(\Delta_{{\bm Q}} \Delta^*_{\bar{{\bm Q}}}\Delta_{-{\bm Q}} \Delta^*_{-{\bar{\bm Q}}}+\text{h.c.}\right) +\ldots
\label{freeenergyexpansion}
\end{align}
A similar phenomenological expression for the free energy (for a system with the $C_4$ symmetry of a square lattice) was given by Agterberg and 
Tsunetsugu. \cite{Agterberg-2008}

Knowing the expression for the coefficients in GL free energy we can see which state is favorable. This is equivalent to computing the GL free
energy for each SC state and compare them to see which one is the lowest. For the FF state we will use that ansatz that the only non zero order 
parameter is $|\Delta_{\bm Q}|$, for the unidirectional PDW we will assume $|\Delta_{{\bm Q}}|=|\Delta_{-{\bm Q}}|$, for the bidirectional 
PDW $|\Delta_{{\bm Q}}|=|\Delta_{-{\bm Q}}|=|\Delta_{{\bm \bar{\bm Q}}}|=|\Delta_{-{\bm \bar{\bm Q}}}|$,
for the double-helix $|\Delta_{{\bm Q}}|=|\Delta_{{\bm \bar{\bm Q}}}|$ and for the bidirectional time-reversal-breaking PDW (Bi2) phase 
$|\Delta_{{\bm Q}}|=|\Delta_{-{\bm Q}}|=|\Delta_{{\bm \bar{\bm Q}}}|=|\Delta_{-{\bm \bar{\bm Q}}}|$. 
Then for each state we have the following SC free energies:
\begin{align}
 F_{\text{FF}}=&\frac{c^{\text{FF}}_2}{2}|\Delta_{{\bm Q}}|^2+\frac{c^{\text{FF}}_4}{4}|\Delta_{{\bm Q}}|^4+\ldots\\
 F_{\text{PDW}}=&\frac{c^{\text{PDW}}_2}{2}|\Delta_{{\bm Q}}|^2+\frac{c^{\text{PDW}}_4}{4}|\Delta_{{\bm Q}}|^4+\ldots \\
 F_{\text{Bi}}=&\frac{c^{\text{Bi}}_2}{2}|\Delta_{{\bm Q}}|^2+\frac{c^{\text{Bi}}_4}{4}|\Delta_{{\bm Q}}|^4+\ldots \\
  F_{\text{2H}}=&\frac{c^{\text{2H}}_2}{2}|\Delta_{{\bm Q}}|^2+\frac{c^{\text{2H}}_4}{4}|\Delta_{{\bm Q}}|^4+\ldots\\
 F_{\text{Bi2}}=&\frac{c^{\text{Bi2}}_2}{2}|\Delta_{{\bm Q}}|^2+\frac{c^{\text{Bi2}}_4}{4}|\Delta_{{\bm Q}}|^4+\ldots
 \end{align}
 where 
 \begin{align}
 c^{\text{PDW}}_2=&2c^{\text{FF}}_2\nonumber\\
 c^{\text{PDW}}_4=&2c^{\text{FF}}_4+u\nonumber\\
 c^{\text{Bi}}_2=&4c^{\text{FF}}_2\nonumber\\
 c^{\text{Bi}}_4=&4c^{\text{FF}}_4+2u+4v_1+2v_2\nonumber\\
  c^{\text{2H}}_2=&2c^{\text{FF}}_2\nonumber\\
 c^{\text{2H}}_4=&2c^{\text{FF}}_4+v_1\nonumber\\
 c^{\text{Bi2}}_2=&4c^{\text{FF}}_2\nonumber\\
 c^{\text{Bi2}}_4=&4c^{\text{FF}}_4+2u+4v_1-2v_2
 \end{align}
where the coefficients are given in Appendix \ref{sec:appendix}.
This expansion is only valid provided $c_4>0$. If $c_4<0$ we need to include higher order terms in the expansion to assure  
thermodynamic stability for large $\Delta_{\bm Q}$. 

Using standard perturbation theory (see, e.g. Refs. [\onlinecite{AGD,schrieffer,sakita,Altland-2010,Radzihovsky-2011}]) the computation of the 
coefficients in the free energy reduces to a computation of a set of Feynman diagrams. An  explicit derivation and form of the 
coefficients $c_2$ and $c_4$ for each of the SC states is given in Appendix \ref{sec:appendix}.

\begin{figure}[hbt]
\subfigure[]{\includegraphics[width=0.5\textwidth]{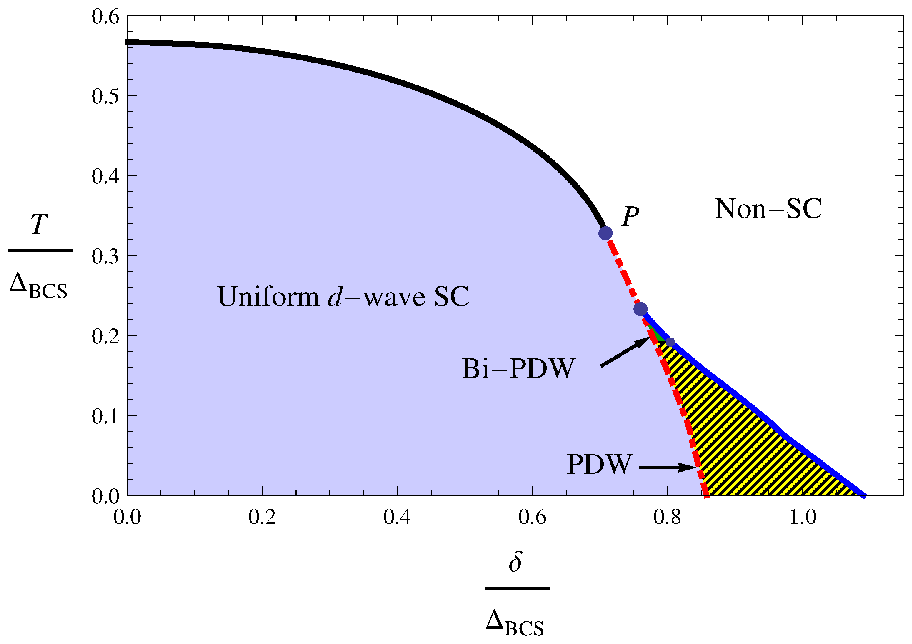}\label{phasediagram}}
\subfigure[]{\includegraphics[width=.18\textwidth]{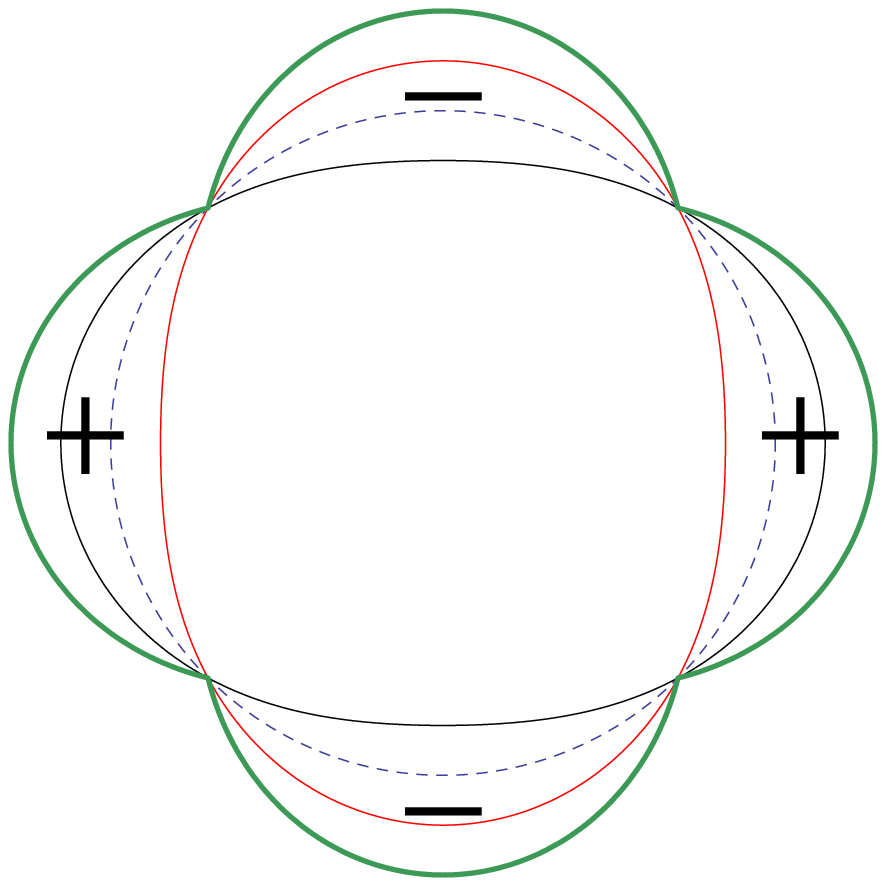}\label{dwavegap}}
\subfigure[]{\put(46,57){${\bm Q}$}\includegraphics[width=.18\textwidth]{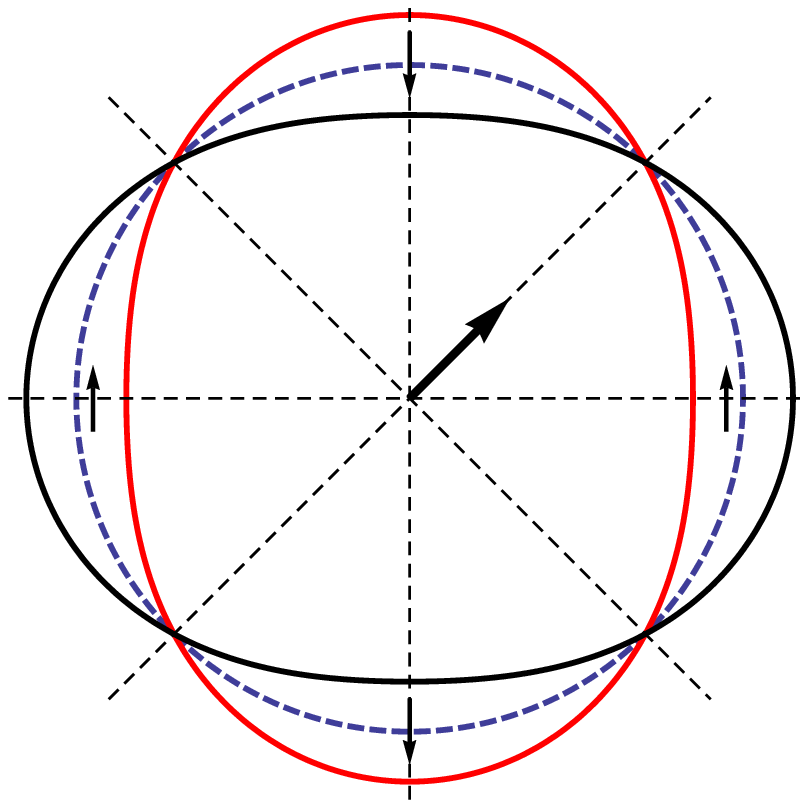}\label{Q1}}\\
\subfigure[]{\put(46,54){${\bm Q}$}\includegraphics[width=.18\textwidth]{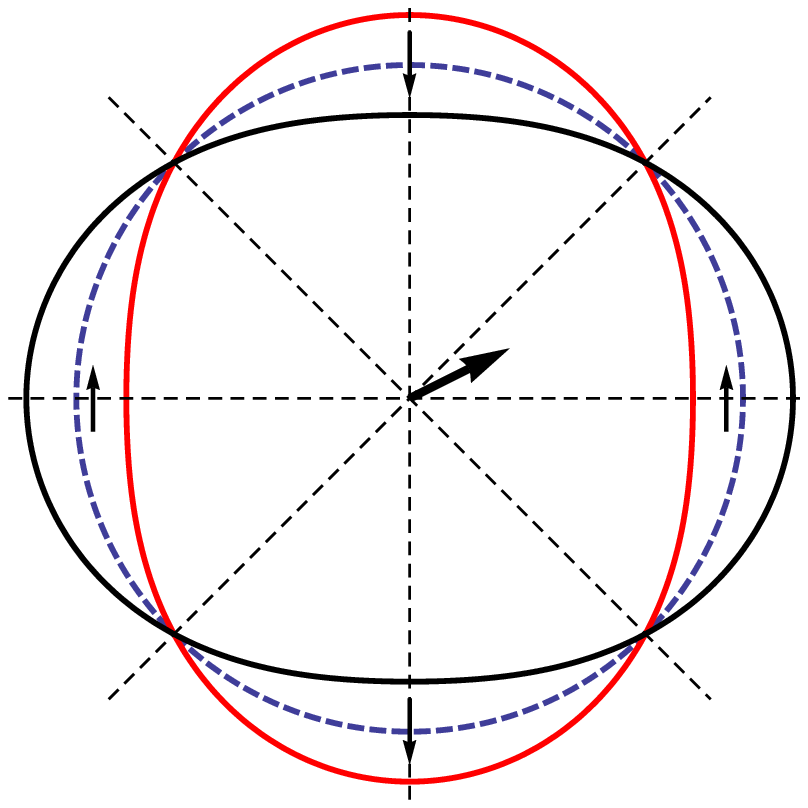}\label{Q2}}
\subfigure[]{\put(50,49){${\bm Q}$}\includegraphics[width=.18\textwidth]{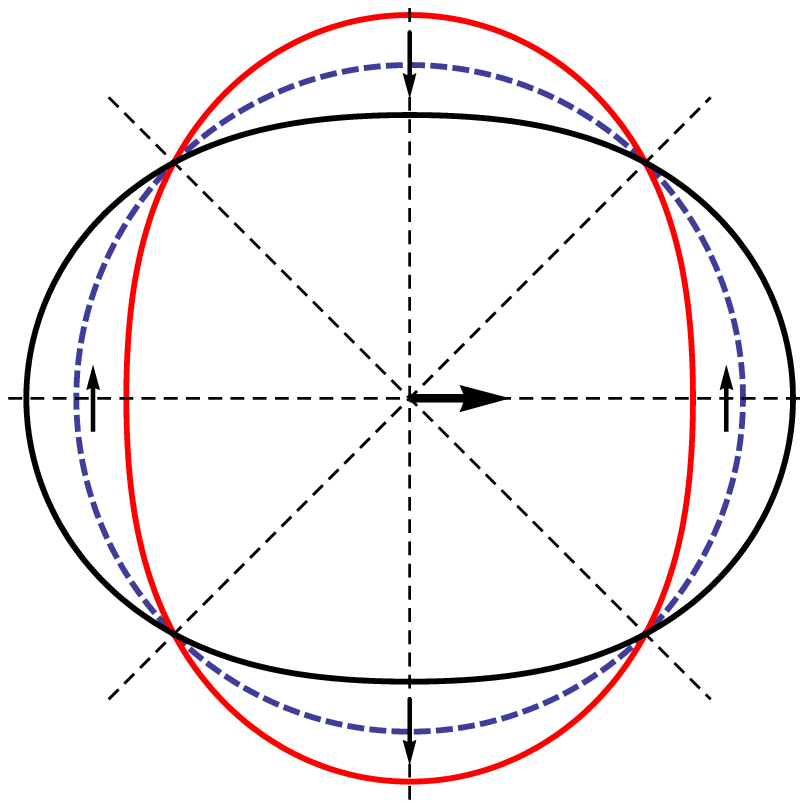}\label{Q3}}
\caption{(color online) (a) Phase diagram as a function of the parameters $T/\Delta_{BCS}$ and $\delta/\Delta_{BCS}$. The light blue region 
corresponds to the uniform $d$-wave SC state (with ${\bm Q}=0$). The yellow and green regions correspond to the PDW and the bidirectional SC 
states respectively. $P$ labels  the tricritical point discussed in the text.(b) Order parameter of the uniform $d$-wave SC state. 
(c)-(e) Direction of the inhomogeneous superconducting  wave vector ${\bm Q}$ 
for (c) $0.23\gtrsim T/\Delta_{BCS}\gtrsim0.055$, (d) $ 0. 055\gtrsim T/\Delta_{BCS}> 0$ and (e) $T=0$. Smooth curves show continuous phase transitions and first order transitions are shown as dashed curves.}
\label{phasediagram2}
\end{figure}

We find that for the range of parameters that we considered $c_4>0$. 
For $c_2>0$ the minimum is at $|\Delta_{{\bm Q}}|=0$, with $F=0$. For $c_2<0$ the minimum is at $|\Delta_{{\bm Q}}|=\sqrt{|c_2|/c_4}$, with
$F=-c_2^2/4c_4$. Computing numerically the coefficients for the FF, double-helix, unidirectional PDW, bidirectional PDW and  
 bidirectional time-reversal-breaking  PDW SC states we then compare their respective free energies resulting in the phase diagram shown in Fig. \ref{phasediagram}. 
 
For $T/\Delta_{BCS}\gtrsim0.33$, and provided the pairing coupling constant $g_\alpha^d$ is larger that its critical value,  
there is a continuous transition from the normal (Non-SC) state to the uniform $d_{x^2-y^2}$-wave SC state, where the 
conventional BCS SC gap $\Delta_{BCS}=2\omega_D\exp(-1/gN_F)$ is introduced to parametrize the dependence on $g$ and $\omega_D$, 
where $\omega_D$ is a high energy cutoff. For the isotropic state, $\delta=0$,
we recover the usual BCS second order transition at $T\approx0.5669\Delta_{BCS}$. 
However, for $0.33\gtrsim T/\Delta_{BCS}\gtrsim0.23$ the transition from the normal (Non-SC) state to the uniform $d$-wave SC state is found to be 
first order, where  there is a tricritical point, $T_{TCP}\simeq 0.33 \Delta_{BCS}$. 

The nodal directions of the $d_{x^2-y^2}$-wave state are, as usual, along the diagonals. In the $\alpha$ phase these directions are 
symmetry directions  where the two Fermi surfaces intersect each other, while the antinodal directions point along the lobes of the Fermi surface 
(see Fig.\ref{dwavegap}). 
A putative $d_{xy}$-wave SC state would have its antinodal directions along the diagonals. However this state is not favored since the isotropic Fermi surface has been effectively gapped (except at a set of zero measure) leading, once again, to a state with a critical coupling constant. In addition, the $d_{xy}$ form factor does not favor inhomogeneous SC states. We will not discuss this channel in what follows.

The most interesting part of the phase diagram is for $T/\Delta_{BCS}\lesssim0.23$. 
In this region there is a continuous transition from the normal (Non-SC) state to an inhomogeneous superconducting  state. Here we find 
two distinct phase 
transitions.
For the temperature range $0.23\gtrsim T/\Delta_{BCS}\gtrsim0.20$ there is a continuous phase transition from the normal (Non-SC) state to a 
bidirectional PDW  state, while for $T/\Delta_{BCS}\lesssim0.20$ there is a continuous transition to a unidirectional  PDW SC state. 
The ordering wave vector for the bidirectional PDW state is locked along the diagonal direction of the spin triplet nematic $\alpha $ phase 
(as shown in Fig. \ref{Q1}). We also find that, for this model, the time-reversal breaking phases (FF, 2H and Bi2) are not energetically favorable,
at least close enough to the transition to the normal state.

In addition, we find a transition from the bidirectional PDW SC to the uniform $d$-wave SC state and from the unidirectional PDW to uniform 
$d$-wave state. Since our expansion for the free energy Eq. 
\eqref{freeenergyexpansion} is only valid close to the continuous transition, we investigate this transition  using the
exact expression for the free energy Eq. \eqref{freenergySC}, and find that this transition is  first order. 
However, the exact expression for the free energy Eq. \eqref{freenergySC} is valid only for the uniform and the FF SC states, 
and hence it can describe only the putative transition from the uniform $d$-wave to the PDW states, depicted by a  dashed curve  
in Fig. \ref{phasediagram}. The actual transition from the PDW SC states to the uniform $d$-wave SC state cannot be described by this free 
energy and it is most likely to occur to the left of the dashed curve. Deeper in the phase diagram the phases that break translation invariance 
should be described by an ansatz that includes many harmonics and, hence, it is best described as a sequence of domain walls, or 
discommensurations, as in the theory of charge-density-waves.

We also investigated the possibility of coexistence of the inhomogeneous superconducting  state and the uniform $d$-wave SC state. We found 
that this does not happen and that the system prefers to be either in the pure inhomogeneous superconducting state or in the pure 
uniform $d$-wave SC state. In addition our results suggest that the continuous phase transition from the normal $\alpha$ phase to the 
bidirectional PDW state merges with the first order transition into the $d$-wave state. This feature is not generic and it is likely  to be  an artifact of the model.

In our analysis we find that the direction (and magnitude) of the ordering wave vector ${\bm Q}$ changes along the continuous phase boundary 
from the normal to the inhomogeneous superconducting  state (as shown in Figs. \ref{Q1}, \ref{Q2}, and \ref{Q3}.)
At $T=0$, ${\bm Q}$ points along the direction of maximum distortion $\phi=n\pi/2$, where $n\in\mathbb{Z}$ and $Q=|{\bm Q}|=2\delta$ (Fig.\ref{Q3}), and hence there are two possible orientations for the unidirectional PDW state. 
As the temperature increases, ${\bm Q}$ rotates continuously towards the diagonal directions (Fig.\ref{Q2}) and for $T/\Delta_{BCS}\gtrsim 0.055$ locks to the diagonal directions 
$\phi=n\pi/2+\pi/4$ (Fig.\ref{Q1}) where, at a somewhat higher temperature, the ordering becomes bidirectional along the two diagonals. In the intermediate regime there are four possible orientations for the unidirectional PDW state which reduce to two directions once the ordering wave vector locks along the diagonal direction of the $\alpha$ phase. We only find bidirectional PDW order along the principal axes of the $\alpha$ phase.  A similar evolution of ordering wave vectors was found in studies  of  2D FFLO phases due to the presence of a Zeeman magnetic field.\cite{Shimahara-1998}

So far we have only considered an attractive pairing interaction in the $d$-wave  channel. However, it is also possible to have spin-triplet 
superconductivity, e.g. $p$-wave, even if the microscopic interactions are nominally repulsive.\cite{Raghu-2010,Kohn-1965,Chubukov-1993}
 In this case we can have pairing between fermions with the same spin polarization (up-up and down-down).
As we can see from Fig. \ref{fig:ellipses} there is perfect nesting, so there is an infinitesimal SC instability in the spin-triplet channel 
(with zero center of mass momentum of the Cooper pairs). This SC state is dominant for small values of the coupling constant. 
However, if the coupling constant in 
the $d$-wave channel is larger than a critical value $g_c$, it will be a competition between the $d$-wave SC state and the spin-triplet SC state.
We considered possible coexistence and competition between both phases ($d$-wave and uniform $p$-wave). We found that there is no coexistence between such
phases and that the state with a larger $T_c$ will be dominant. We can effectively tune the coupling constant in the $d$-wave channel in order make the
$d$-wave SC state favorable against the $p$-wave SC state (or, equivalently, lower the coupling constant in the $p$-wave channel in order to decrease
its $T_c$).

\subsection{$\beta$-phase MF}
\label{sec:beta-MF}

\begin{figure}[hbt]
\begin{center}
\includegraphics[width=0.5\textwidth]{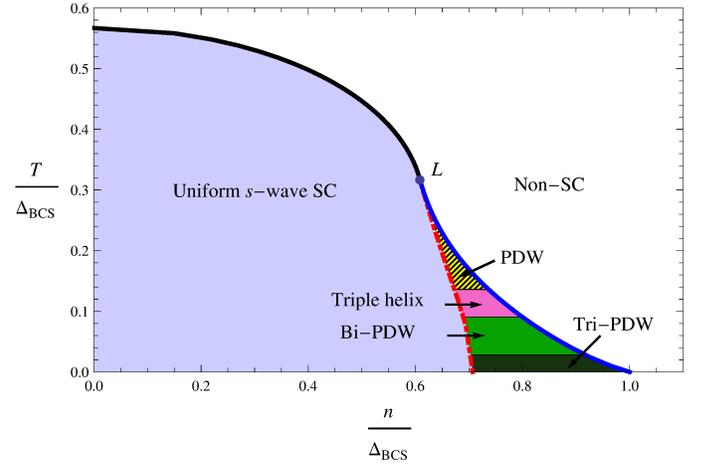}
\end{center}
\caption{(color online) Phase diagram in the nematic $\beta$-phase for $s$-wave pairing. Here $n$ is the magnitude of the order parameter of the spin triplet nematic $\beta$ phase. All transitions from the non-SC $\beta$ phase are continuous. Notice the complex sequence of phases with different types of homogeneous and inhomogeneous orders, and their sequence of multicritical points. Here $L$ labels the Lifshitz point discussed in the text. Smooth curves show continuous phase transitions and first order transitions are shown as dashed curves. The same caveats we pointed out in the case of the  $\alpha$ phase apply here too. See text for details. }
\label{fig:beta-diagram-swave}
\end{figure}

\begin{figure}[hbt]
\begin{center}
\includegraphics[width=0.5\textwidth]{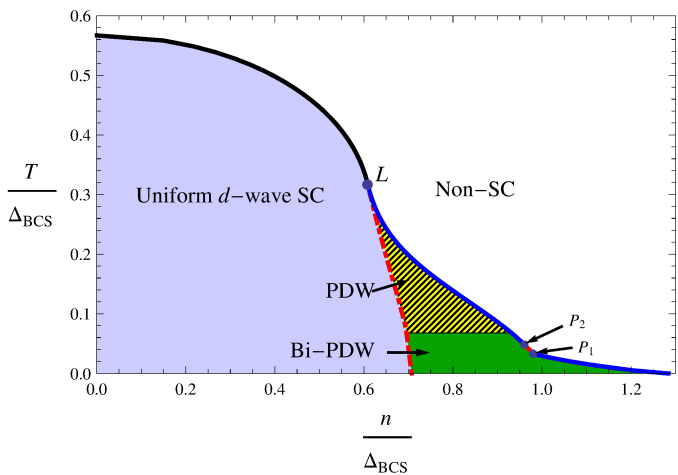}
\end{center}
\caption{(color online) Phase diagram in the nematic $\beta$-phase for $d$-wave pairing. Here $n$ is the magnitude of the order parameter of 
the spin-triplet nematic $\beta$ phase. $L$ is the Lifshitz point separating the uniform d-wave SC from the unidirectional PDW phase. $P_2$ is a 
tricritical point separating a continuous non-SC high temperature phase from the bidirectional PDW phase, and $P_1$ is a bicritical point 
separating the low temperature bidirectional phase oriented along the anti-nodal direction of the $d$-wave from a bidirectional PDW oriented along 
the nodal direction. Smooth curves show continuous phase transitions and first order transitions are shown as dashed curves. See text for details. }
\label{fig:beta-diagram-dwave}
\end{figure}

We now turn to the case of the nematic triplet $\beta$ phase and look for the possible superconducting states that may occur.
For the choice $|{\bm n}_1|=|{\bm n}_2|=\bar{n}$ and ${\bm n}_1=\bar{n}\hat{x}$ 
and  ${\bm n}_2=\bar{n}\hat{y}$ the Hamiltonian in the $\beta$ phase can be written as:
\begin{align}
 H=&\sum_{{\bm k},\alpha,\beta}c^{\dagger}_{{\bm k},\alpha}(\epsilon_{{\bm k}}-\bar{n}{\bm d}_{{\bm k}}
\cdot{\bm \sigma}_{\alpha,\beta})c^{}_{{\bm k},\beta}\nonumber \\
&+\sum_{{\bm k},{\bm k}',{\bm q}}V({\bm k},{\bm k}')c^{\dagger}_{{\bm k}+{\bm q}/2,\uparrow}c^{\dagger}_{-{\bm k}+{\bm q}/2,\downarrow}
c^{}_{-{\bm k}'+{\bm q}/2,\downarrow}c^{}_{{\bm k}'+{\bm q}/2,\uparrow}
 \label{Hamiltonianbeta}
\end{align}
where ${\bm d}_{{\bm k}}=(\cos(l\theta_k),\sin(l\theta_k),0)$.
\begin{equation}
{\bm d}_{{\bm k}}\cdot{\bm \sigma}=
\displaystyle\left( \begin{array}{cc}
0 & e^{-il\theta_k} \\
e^{il\theta_k} & 0 \end{array} \right)
\label{SOC}
\end{equation}
As we saw
in the Section \ref{sec:SCinst}, for the $d$-wave channel  there are four equivalent directions for which the SC susceptibility for the 
$\beta$-phase has a maximum. Thus, as for the $\alpha$-phase, we can focus on different cases for the superconducting order parameters: 
FF, double-helix, PDW, bidirectional PDW and time-reversal breaking PDW. 
However, for $s$-wave pairing  all the directions are equivalent, allowing us to have in principle orderings in all possible directions.
Nevertheless, we will only study in addition to the FF, PDW and bidirectional PDW SC states, the tridirectional PDW and the triple helix state, 
which are expected to be favored on the basis of symmetry. These phases are defined as the follows:
\begin{itemize}
\item Triple helix phase: in this phase three wave vectors contribute to the SC order parameter,
\begin{equation}
\Delta({\bm r})=\Delta_{{\bm Q_1}}e^{i{\bm Q_1}\cdot{\bm r}}+\Delta_{{\bm Q_2}}e^{i{\bm Q_2}\cdot{\bm r}}+
\Delta_{{\bm Q_3}}e^{i{\bm Q_3}\cdot{\bm r}}
\label{triplehelix1}
\end{equation}
where the angle between the ${\bm Q_i}$'s is $2\pi/3$. Assuming from now on $|\bm Q_1|=|\bm Q_2|=|\bm Q_3|$, so that 
${\bm Q}_1+{\bm Q}_2+{\bm Q}_3=0$, $|\Delta_{{\bm Q_1}}|=|\Delta_{{\bm Q_2}}|=|\Delta_{{\bm Q_3}}|$,
and neglecting the phase fluctuations of these three complex order parameters, we write the previous expression as:
\begin{equation}
 \Delta({\bm r})=\Delta_{|{\bm Q_1}|}\left( e^{i{\bm Q_1}\cdot{\bm r}}+e^{i{\bm Q_2}\cdot{\bm r}}+e^{i{\bm Q_3}\cdot{\bm r}} \right)
\label{triplehelix}
\end{equation}
\item Tridirectional PDW phase: in this phase six wave vectors contribute to the SC order parameter,
\begin{align}\nonumber
\Delta({\bm r})=&\Delta_{{\bm Q_1}}e^{i{\bm Q_1}\cdot{\bm r}}+\Delta_{-{\bm Q_1}}e^{-i{\bm Q_1}\cdot{\bm r}}\\ \nonumber
&+\Delta_{{\bm Q_2}}e^{i{\bm Q_2}\cdot{\bm r}}+\Delta_{-{\bm Q_2}}e^{-i{\bm Q_2}\cdot{\bm r}}\\
&+\Delta_{{\bm Q_3}}e^{i{\bm Q_3}\cdot{\bm r}}+\Delta_{-{\bm Q_3}}e^{-i{\bm Q_3}\cdot{\bm r}}
\label{tridirectional1}
\end{align}
In this phase the SC order parameter is then a six-component complex field with $\Delta_{\pm {\bm Q_i}}$, where $i=1,2,3$, being the six 
complex components (and hence six amplitudes and six phase fields). Under the assumption of parity and $C_{6}$ symmetry it reduces to
\begin{equation}
 \Delta({\bm r})=2|\Delta_{{\bm Q}}|\left( \cos({\bm Q_1}\cdot{\bm r})+\cos({\bm Q_2}\cdot{\bm r})+\cos({\bm Q_3}\cdot{\bm r}) \right)
\label{tridirectional}
\end{equation}
where we  assumed that the tree ordering wave vector have the same magnitude, $|{\bm Q}_1|=|{\bm Q}_2|={\bm Q}_3|=|{\bm Q}|$ and that the angle 
between these vectors is $2\pi/3$. In addition we also assumed that 
$|\Delta_{{\bm Q}_1}|=|\Delta_{-{\bm Q}_1}|=|\Delta_{{\bm Q}_2}|=|\Delta_{-{\bm Q}_2}|=|\Delta_{{\bm Q}_3}|=|\Delta_{-{\bm Q}_3}|$
\end{itemize}
Since the possible SC phases for the $\beta$ is larger than what we found in the case of the $\alpha$ phase, the associated Landau free energy 
has a more complex for. We will not exhibit it here in its full form (for a general phenomenological expression for the free energy
see for instance Ref. [\onlinecite{Agterberg-2011}]).
We will compute the free energy for each one of these phases in order to determine the phase diagram as we did for the $\alpha$-phase.
As for the $\alpha$-phase we can perform a Hubbard-Stratonovich transformation to decouple the  pairing interactions. 
Here too the calculation simplifies for the FF phases since the Green function matrix is block diagonal, with each block being labeled by the momentum ${\bm k}$ and the Matsubara frequency $\omega_n$. Also, as what we found in the $\alpha$ phase, the free energy in the PDW SC states cannot be computed in closed form and can be obtained as a power series expansion in the PDW order parameters, whose coefficients need to be evaluated numerically. This analysis leads to the phase diagrams shown in  Fig.\ref{fig:beta-diagram-swave} and Fig. \ref{fig:beta-diagram-dwave}.

The action for the FF state  is
\begin{align}
 S[\bar{\Psi},\Psi,\Delta_{{\bm Q}},\Delta^*_{{\bm Q}}]&=\nonumber\\
-\sum_{{\bm k},n}\bar{\Psi}_{{\bm k},n}\mathcal{G}^{-1}_{{\bm k},i\omega_n}&\Psi_{{\bm k},n}+\beta\frac{|\Delta_{{\bm Q}}|^2}{g}
+\text{const.},
\end{align} 
where now the Nambu operator $\bar{\Psi}_{{\bm k}}$  is given by
\begin{equation}
\bar{\Psi}_{{\bm k}}=(\bar{\psi}_{{\bm k}+{\bm Q}/2,\uparrow},\bar{\psi}_{{\bm k}+{\bm Q}/2,\downarrow},
\psi_{-{\bm k}+{\bm Q}/2,\uparrow},\psi^{}_{-{\bm k}+{\bm Q}/2,\downarrow})
\end{equation}
The inverse of the Green function for fixed ${\bm k}$ and $\omega_n$, $\mathcal{G}_{{\bm k},i\omega_n}$, is given by the $4\times 4$ matrix
\begin{widetext}
\begin{equation}
\mathcal{G}^{-1}_{{\bm k},i\omega_n}=\frac{1}{2}
\displaystyle\left( \begin{array}{cccc}
i\omega_n-\xi_{{\bm k}+{\bm Q}/2,\uparrow} & ne^{-il\theta_{{\bm k}+{\bm Q}/2}} & 0 &\Delta_{{\bm Q}}\gamma({\bm k}) \\
ne^{il\theta_{{\bm k}+{\bm Q}/2}} & i\omega_n-\xi_{{\bm k}+{\bm Q}/2,\downarrow} & -\Delta^*_{{\bm Q}}\gamma({\bm k}) & 0\\
0 & -\Delta_{{\bm Q}}\gamma({\bm k}) & i\omega_n+\xi_{-{\bm k}+{\bm Q}/2,\uparrow} & -ne^{il\theta_{-{\bm k}+{\bm Q}/2}}\\ 
\Delta^*_{{\bm Q}}\gamma({\bm k}) & 0 & -ne^{-il\theta_{-{\bm k}+{\bm Q}/2}} & i\omega_n+\xi_{-{\bm k}+{\bm Q}/2,\downarrow} 
\end{array} \right)
\label{Greenbeta}
\end{equation}
\end{widetext}
In the $\beta$-phase $\xi_{{\bm k},\uparrow}=\xi_{{\bm k},\downarrow}=\xi_{{\bm k}}$ and $\xi_{{\bm k}}=\xi_{{\bm -k}}$.
Also notice that $\theta_{-{\bm k}}=\theta_{{\bm k}}+\pi$. 

After integrating out the fermionic degrees of freedom we get:
\begin{align}
 S_{\text{eff}}[\Delta^*_{{\bm Q}},\Delta_{{\bm Q}}]=&-\sum_{{\bm k},n} \ln \det \mathcal{G}_{{\bm k}, n}^{-1}+\beta\frac{|\Delta_{{\bm Q}}|^2}{g}+\text{const.}\nonumber\\
 =&-\sum_{{\bm k}, n} \sum_{j=1}^4 \ln \lambda_{{\bm k},n}^{(j)}+\beta\frac{|\Delta_{{\bm Q}}|^2}{g}+\text{const.}
\label{effectiveactionbeta}
\end{align}
 where 
$\lambda^{(j)}_{{\bm k},n}$ are the eigenvalues of $\mathcal{G}^{-1}_{{\bm k},i\omega_n}$.
Using that 
\begin{equation}
\xi_{\pm {\bm k}+{\bm Q}/2}=\xi \pm Q/2\cos(\theta-\phi)
\end{equation}
we find that:
\begin{align}
 \lambda^{(j)}_{{\bm k},n}=&E_j-i\omega_n
\label{evsbeta}
\end{align}
where $E_j$ are the eigenvalues of the $4\times 4$ matrix $2\mathcal{G}^{-1}_{{\bm k},i\omega_n}-i\omega_n I$. 

Since we are interested in the inhomogeneous superconducting  state, we focus from 
now on $\beta$ phases with even $l$. 
The Free energy in the $\beta$-phase is given by
\begin{align}
 F_s-F_n=&\frac{|\Delta_{{\bm Q}}|^2}{g}
 \nonumber\\
 -TN(E_F) &\; \int_{0}^{2\pi}\frac{d\theta}{2\pi}\int_{0}^{\omega_D}d\xi\sum_{j=1}^4\ln \left(\frac{1+e^{-E_j(\Delta_{{\bm Q}})/T}}
{1+e^{-E_j(\Delta=0)/T}}\right)
\label{freenergySCbeta}
\end{align}
Once again, we now  minimize the free energy with respect to $\Delta$ and ${\bm Q}$ to find the equilibrium state.
In contrast with the $\alpha$ phase, in the $\beta$-phase is not necessary to have $d$-wave pairing to have an inhomogeneous superconducting state. Thus, we can now have $s$- or $d$-wave pairing. We will study below
the phase diagram for the $\beta$-phase for both SC channels.

The resulting phase diagrams have a rich structure.
We find again that the transition from the normal (non-SC state) to the inhomogeneous superconducting  state is continuous. 
However, in contrast to the $\alpha$-phase case, we found that transition is continuous at all temperatures even as the inhomogeneous states 
meet the uniform states ($s$ or $d$ wave depending on the case). Thus in the $\beta$ phase the transition from the normal (Non-SC) state 
to the uniform SC state and to the unidirectional PDW SC state is continuous and are shown in the phase diagrams of 
Fig. \ref{fig:beta-diagram-swave} and Fig. \ref{fig:beta-diagram-dwave} respectively.

In particular the transitions between the uniform SC states and the unidirectional PDW states are multicritical points which have the same 
structure as the well known Lifshitz points of magnetism and liquid crystals.\cite{chaikin-1995} Near the Lifshitz points the SC susceptibilities 
can be expanded in powers of the magnitude of the ordering wave vector $Q=|{\bm Q}|$ in the form
\begin{align}
 \frac{\chi_{\beta}^{\lambda}({\bm Q})}{N(E_F)}=\chi_0+\chi_2 Q^2+\chi_4 Q^4 + O(Q^6)
 \label{suscbetaexp}
 \end{align}
 where the coefficients $\chi_0$, $\chi_2$ and $\chi_4$ need to be computed numerically. The important feature of the SC susceptibility is that the coefficient $\chi_2$ continuously changes along the phase boundary between the normal state and the uniform SC state 
from  positive  to negative values across the Lifshitz point where it vanishes. The other two coefficients, $\chi_0$ and $\chi_4$, also vary smoothly but without changing sign.

As we did for the $\alpha$-phase we can compute the Ginzburg-Landau free energy for the $\beta$-phase. We need again to determine the 
coefficients $c_2$ and $c_4$ in the GL free energy for the different SC phases. This can be done in the same way we determined the 
GL free energy coefficients for the $\alpha$-phase (explained in detail in the appendix \ref{sec:appendix}).
In addition, as it was mentioned above, there is a parallel in the calculations for the expression with even $l$ in the $\beta$-phase and 
the calculations carried by Shimahara in Ref. \onlinecite{Shimahara-1994}. Therefore we can use his results to determine the phase diagrams
for the $s$- and $d$-wave  channels.

For the $s$-wave pairing we found the following. For $0.318\gtrsim T/\Delta_{BCS}\gtrsim0.136$ the favored SC state is a unidirectional PDW 
state with wave vector ${\bm Q}$ with arbitrary direction. The unidirectional PDW SC phase meets the uniform $s$-wave SC at the Lifshitz point 
shown in Fig.\ref{fig:beta-diagram-swave}. The wave vector of the unidirectional PDW state grows continuously from zero away from the Lifshitz 
point.\cite{chaikin-1995}  In the temperature range $0.136\gtrsim T/\Delta_{BCS}\gtrsim0.091$ a triple helix FF state is favorable, 
while for  $0.091\gtrsim T/\Delta_{BCS}\gtrsim0.028$ the bidirectional PDW state is favorable. Finally, 
for $T/\Delta_{BCS}\lesssim0.028$ a tridirectional PDW state is the favored. These results are summarized in 
the phase diagram in Fig. \ref{fig:beta-diagram-swave}.

It was pointed out in Ref. [\onlinecite{Agterberg-2011}] in the Tri-PDW state is  possible to have additional terms of the free energy of the form 
$\Delta_{{\bm Q}_1}\Delta_{{\bm Q}_2}\Delta^*_{-{\bm Q}_3}\Delta^*_{0}$ in the GL free energy, where $\Delta^*_{0}$ is the order parameter 
for the uniform s-wave. Such terms arise if there is a coexistence phase between the tridirectional PDW state and the uniform $s$-wave SC state. Similarly, higher charge subdominant uniform SC states are expected to occur in in PDW phases.\cite{Berg-2009,Berg-2009b}

In the case of $d$-wave pairing the situation is different. The phase diagram for the $\beta$ phase in the $d$-wave case  is shown in Fig. \ref{fig:beta-diagram-dwave}.
Here we find that  the direction of ordering wave vector $\vec{Q}$ of the inhomogeneous SC states  is no longer arbitrary and it is not same  throughout these superconducting states. For
$T/\Delta_{BCS}\lesssim0.034$, the ordering wave vector ${\bm Q}$ points in the antinodal directions of the $d$-wave with $\phi=n\pi/2$, where $n\in\mathbb{Z}$. 
On the other hand, for $T/\Delta_{BCS}\gtrsim0.034$, the ordering wave vector  ${\bm Q}$ now points in the nodal directions with $\phi=n\pi/2+\pi/4$. 

For $T/\Delta_{BCS}\lesssim0.034$
we find a bidirectional PDW state whose ordering wave vector $\bm Q$  points along the antinodal directions of the $d$-wave.
For $0.034\lesssim T/\Delta_{BCS} < 0.068$ we find a bidirectional PDW 
state whose ordering wave vector $\bm Q$ points along the nodal directions of the $d$-wave. This means that 
above the tricritical point $P_1$ shown in Fig.\ref{fig:beta-diagram-dwave} (where $T/\Delta_{BCS}=0.034$) the direction of the ordering wave 
vector $\bm Q$ rotates by $\pi/4$ and above $P_1$ points along the nodal directions of the $d$-wave. 
Above the tricritical point $P_1$ the transition from the normal state (Non-SC) to the bidirectional PDW state is first order and becomes 
continuous at a the second tricritical point $P_2$ at $T/\Delta_{BCS}=0.048$. 
The coefficient $c_4$ of the Landau free energy vanishes at both tricritical points (as it should). 
For $0.068\lesssim T/\Delta_{BCS}\lesssim 0.318$ we find a unidirectional PDW state whose ordering wave vector $\bm Q$ 
points along the nodal directions of the $d$-wave.

\section{Concluding Remarks}
\label{sec:conclusions}

A principal motivation for this work was to investigate using controlled approximations the possible relation between electronic liquid crystal phases (of which the spin triplet nematic states are just two examples) and superconductivity. Our results indicate that this type of electronic liquid crystal phases naturally give raise to complex inhomogeneous superconducting phases. 
Unfortunately, so far as we know, spin-triplet nematic metallic phases have yet to be discovered in experiment. 

Earlier studies of superconducting instabilities in spin-singlet nematic phases did not reach a clear answer.\cite{kee-2004,yamase-2007b} In addition, PDW phases and others of similar nature, are notoriously difficult to study as they are outside the reach of weak coupling BCS theory.\cite{Loder-2010,Loder-2011} On the other hand,  high-quality numerical tensor-network approaches such as iPEPS have provided solid evidence for the existence (or, at least, competitiveness) of PDW phases in simple 2D strongly correlated systems such as the $t-J$ model.\cite{corboz-2014} Interestingly these authors find that some degree of fixed nematicity (i.e. explicit rotational symmetry breaking) strongly favors PDW ordered SC states. In addition, commensurate PDW phases have been shown to occur in Kondo-Heisenberg chains\cite{Berg-2010} and in doped spin-ladders.\cite{Jaefari-2012} 

The models we studied is this paper are, on the other hand, too idealized as they stand to be relevant to the physics of the cuprates. In addition of the important role of magnetism that these spin-triplet phases imply, for which there is no evidence in these materials, we have used a continuum description  with simple (nearly circular) Fermi surfaces. In strongly correlated systems  rotational spatial symmetry is strongly broken down to the point group symmetry of the lattice. So far there are few studies of lattice models with nematic spin-triplet phases\cite{fischer-2011,Maharaj-2013} and they typically find that this Pomeranchuk type phase transition requires a substantial value of the one-site Hubbard  interactions which put them outside the regime in which their mean field theories may be reliable. Nevertheless these results are interesting and suggest that PDW type phases may also arise in these models. 

Using weak coupling BCS-type methods we showed that nematic spin triplet $\alpha$ and $\beta$ phases give rise to a complex phase diagram which includes pair-density-wave phases and other spatially inhomogeneous superconducting states.  Rather than considering a specific microscopic model we used instead effective pairing interactions in different channels ($s$, $p$ and $d$ wave), with effective coupling constants for each,  and investigated what superconducting states arose as instabilities of the $\alpha$ and $\beta$ spin-triplet nematic states. The theory is well controlled by tuning to the nematic spin triplet to normal Fermi liquid  quantum phase transition. The distance to this quantum critical point inside the nematic spin triplet states plays the role of the small parameter which justifies the use of weak coupling mean field theory (BCS) to describe the resulting superconducting states. In this sense, the nematic spin triplet quantum phase transition can be regarded as a complex multicritical point.

An important feature of the phase diagrams that we present here is that the critical temperatures of all the phases have comparable magnitude. This is the consequence of having fine-tuned to the spin-triplet nematic quantum critical point. A puzzling feature of the intertwined orders seen in the experiments in the cuprate superconductors is that the critical temperatures have the same typical magnitude over a substantial range of doping and for rather different materials. It is unreasonable to think that all the cuprate superconductors have conspired to be  fine-tuned to a multicritical point as in the calculation that we have done here. Rather this is presumably a consequence of strong correlation physics as in the recent work of Corboz, Rice and Troyer\cite{corboz-2014}.

Finally, in this paper we assumed that we were deep enough in the spin-triplet nematic phase that its quantum fluctuations can be neglected. This is clearly not the case close enough to the quantum phase transition from the Fermi liquid phase. In addition we also neglected the possible role of Goldstone modes of the nematic triplet state. These modes, which are gapped by lattice effects, may drive the fermionic fluid into a non-Fermi liquid regime in their absence and change the physics of the superconducting state in an essential way. In particular, in the absence of lattice effects, the Goldstone modes may invalidate the use of BCS-type schemes which require the existence of sharply defined quasiparticles.

\begin{acknowledgments}

We would like to thank S. Raghu, N. Tubman, A. Kogar, S. Kivelson and M. Stone for useful discussions.
This work was supported in part by the National Science Foundation, under grant DMR-1064319  at the University of Illinois, by the U.S. Department of Energy, Division of Materials Sciences under Award No. 
DE-FG02-07ER46453 through the Frederick
Seitz Materials Research Laboratory of the University of Illinois at Urbana-Champaign, and the Program Becas Chile (CONICYT) (RSG).
\end{acknowledgments}

\appendix
\begin{widetext}
\section{Coefficients in the Ginzburg Landau Free Energy}
\label{sec:appendix}
In the following we give a detailed derivation for the coefficients in the GL free energy \eqref{freeenergyexpansion} for the $\alpha$-phase. 
Similar analysis can be carried out for the $\beta$-phase.
Following Radzihovsky \cite{Radzihovsky-2011} (and references therein), we write 
$S=S_0+S_{\text{int}}+\displaystyle\beta\sum_{{\bm Q}}\frac{|\Delta_{{\bm Q}}|^2}{g}$, where:
\begin{align}
\begin{split}
  S_0=&\int_0^{\beta}d\tau\sum_{{\bm k},\sigma}\bar{\psi}_{{\bm k},\sigma}(\partial_{\tau}+\xi_{{\bm k},\sigma})\psi_{{\bm k},\sigma}\\
S_{\text{int}}=&-\int_0^{\beta}d\tau\sum_{{\bm Q}}\sum_{{\bm k}}\left[
\gamma({\bm k})\bar{\psi}_{{\bm k}+{\bm Q}/2,\uparrow}\bar{\psi}_{-{\bm k}+{\bm Q}/2,\downarrow}\Delta_{{\bm Q}}
+\gamma({\bm k})\Delta^*_{{\bm Q}}\psi_{-{\bm k}+{\bm Q}/2,\downarrow}\psi_{{\bm k}+{\bm Q}/2,\uparrow} 
\right]
\end{split}
\end{align}  
The effective action is then given by:
\begin{equation}
 e^{-S_{\text{eff}}}= \int D\psi D\bar{\psi}e^{-S}=e^{-\beta\sum_{{\bm Q}}\frac{|\Delta_{{\bm Q}}|^2}{g}}
 \int D\psi D\bar{\psi}e^{-S_0}e^{-S_{\text{int}}} 
\end{equation}
We can then expand in powers of $S_{\text{int}}$ we obtain
\begin{align}
 \int D\psi D\bar{\psi}e^{-S_0}e^{-S_{\text{int}}} =&
Z_0 \times \left[1+\frac{1}{2!}\langle S^2_{\text{int}}\rangle_0+\frac{1}{4!}\langle S^4_{\text{int}}\rangle_0
+\cdots\right]\nonumber\\
=&Z_0 \times \exp\left(\frac{1}{2!}\langle S^2_{\text{int}}\rangle^c_0+\frac{1}{4!}\langle S^4_{\text{int}}\rangle^c_0+\ldots\right)
\end{align}
where we denoted by $\langle A\rangle_0^c$ the connected expectation value in the normal state and where we used the notation
\begin{equation}
 Z_0\equiv\int D\psi D\bar{\psi}e^{-S_0}, \quad \langle\cdots\rangle_0\equiv \frac{1}{Z_0}\int D\psi D\bar{\psi}e^{-S_0}(\cdots)
 \end{equation}
We also used the fact that the expectation value of odd powers of the interacting part of the action vanishes in the normal state, $\langle S_{\rm int}^{2p+1}\rangle_0=0$.
 
We will now  focus in the quadratic and quartic terms for the FF, the unidirectional PDW, the bidirectional PDW, the double-helix and the 
time-reversal breaking bidirectional PDW SC states:
\begin{enumerate}
 \item FF state.
\begin{enumerate}[label=\emph{\roman*})] 
\item Quadratic term $c^{\text{FF}}_2$.\\
For the quadratic term we need to compute $\displaystyle\frac{1}{2!}\langle S^2_{\text{int}}\rangle_0$, which can be represented by the Feynman 
diagram shown in Fig. \ref{fig:FF-SC-susceptibility-diagram}.
\begin{figure}[hbt]
\begin{center}
\includegraphics[width=0.3\textwidth]{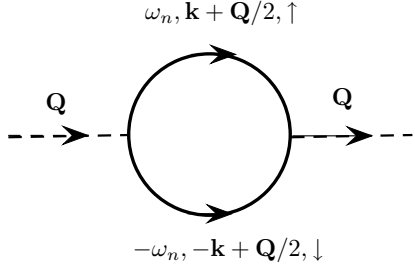}
\end{center}
\caption{Superconducting susceptibility for the FF state.}
\label{fig:FF-SC-susceptibility-diagram}
\end{figure}
This diagram corresponds to the superconducting susceptibility. Adding the term 
$\frac{|\Delta_{{\bm Q}}|^2}{g}$ to the effective action 
and using that $\displaystyle\frac{1}{gN(E_F)}=\ln\left(\frac{2\omega_D}{\Delta_{BCS}} \right)$ we can write $c^{\text{FF}}_2$ as:
\begin{equation}
\frac{c^{\text{FF}}_2}{N(E_F)}=-2\ln\left(\frac{1}{4\pi T} \right)+
2\text{Re}\int_0^{2\pi}\frac{d\theta}{2\pi}2\cos^2(2\theta) \psi\left(\frac12+i\frac{\delta}{2\pi T}\cos(2\theta)-
i\frac{Q/2}{2\pi T}\cos(\theta-\phi)\right)
\label{c2FF}
\end{equation}
where $T$, $Q$ and $\delta$ are in units of $\Delta_{BCS}$, and $\psi(z)$ is the  digamma function (for $z \in \mathbb{C}$)
\begin{equation}
\psi(z)=\frac{d}{dz} \ln \Gamma(z)
\end{equation}
where
\begin{equation}
\Gamma(z)=\int_0^\infty dt \; t^{z-1\; }e^{-t}
\end{equation}
is the Euler Gamma function.
Below we use the standard notation $\psi^{(n)}(z)$ for the derivatives of the digamma function.

\item Quartic term $c^{\text{FF}}_4$.\\
For the quartic term we need to compute 
\begin{equation}
\frac{1}{4!}\langle S^4_{\text{int}}\rangle_0^{c}=\frac{1}{4!}(\langle S^4_{\text{int}}\rangle_0-3\langle S^2_{\text{int}}\rangle_0^2)
\end{equation}
which can be represented by the diagram of Fig. \ref{fig:c4FF-diagram}.
\begin{figure}[hbt]
\begin{center}
\includegraphics[width=0.4\textwidth]{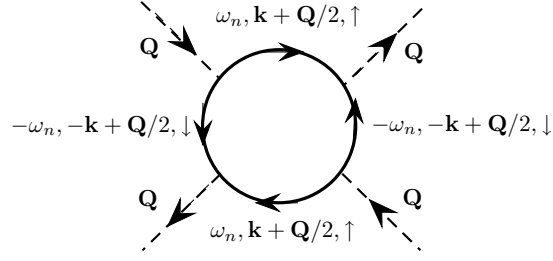}
\end{center}
\caption{Quartic effective interaction for the FF state.}
\label{fig:c4FF-diagram}
\end{figure}
The algebraic expression for this diagram is given by:

\begin{equation}
\frac{c^{\text{FF}}_4}{N(E_F)}=-\frac{1}{8\pi^2 T^2}\int_{0}^{2\pi}\frac{d\theta}{2\pi}4\cos^4(2\theta) 
\text{Re}\left[\psi ^{(2)}\left(\frac12+i\frac{\delta}{2\pi T}\cos2\theta-i\frac{Q/2}{2\pi T}\cos(\theta-\phi)\right)\right]
\label{c4FF}
\end{equation}

\end{enumerate}

\item Unidirectional PDW state.

\begin{enumerate}[label=\emph{\roman*})] 
\item Quadratic term $c^{\text{PDW}}_2$.\\
For the quadratic term we need to compute $\displaystyle\frac{1}{2!}\langle S^2_{\text{int}}\rangle_0$, which can be represented by the 
diagrams of Fig. \ref{fig:LO-SC-susceptibility-diagram}.
\begin{figure}[hbt]
\begin{center}
\includegraphics[width=0.4\textwidth]{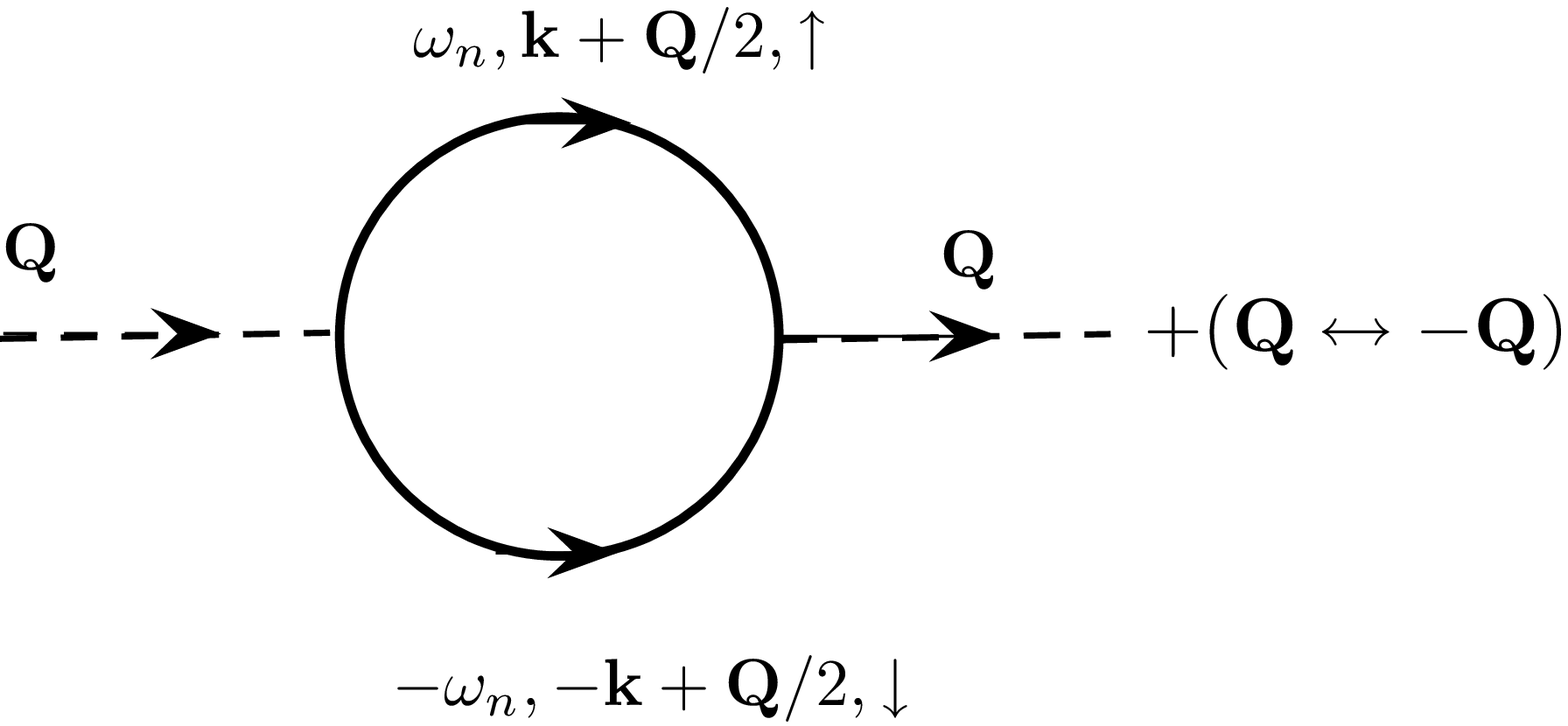}
\end{center}
\caption{Superconducting susceptibility for the LO state.}
\label{fig:LO-SC-susceptibility-diagram}
\end{figure}
Each diagram produce the same contribution. Now remember than in the PDW state we have two plane waves with wave vectors ${\bm Q}$ and 
$-{\bm Q}$, so $\sum_{{\bm q}}\frac{|\Delta_{{\bm q}}|^2}{g}=2\frac{|\Delta_{{\bm Q}}|^2}{g}$ where we used that 
$|\Delta_{{\bm Q}}|=|\Delta_{-{\bm Q}}|$, so we have that
\begin{equation}
\frac{c^{\text{PDW}}_2}{N(E_F)}=2\frac{c^{\text{FF}}_2}{N(E_F)}
\label{c2LO}
\end{equation}
where again $T$, $Q$ and $\delta$ are in units of $\Delta_{BCS}$

\item Quartic term $c^{\text{PDW}}_4$.\\
For the quadratic term we need the diagrams in Fig. \ref{fig:c4PDW}.
\begin{figure}[hbt]
\begin{center}
\subfigure[]{\includegraphics[width=0.4\textwidth]{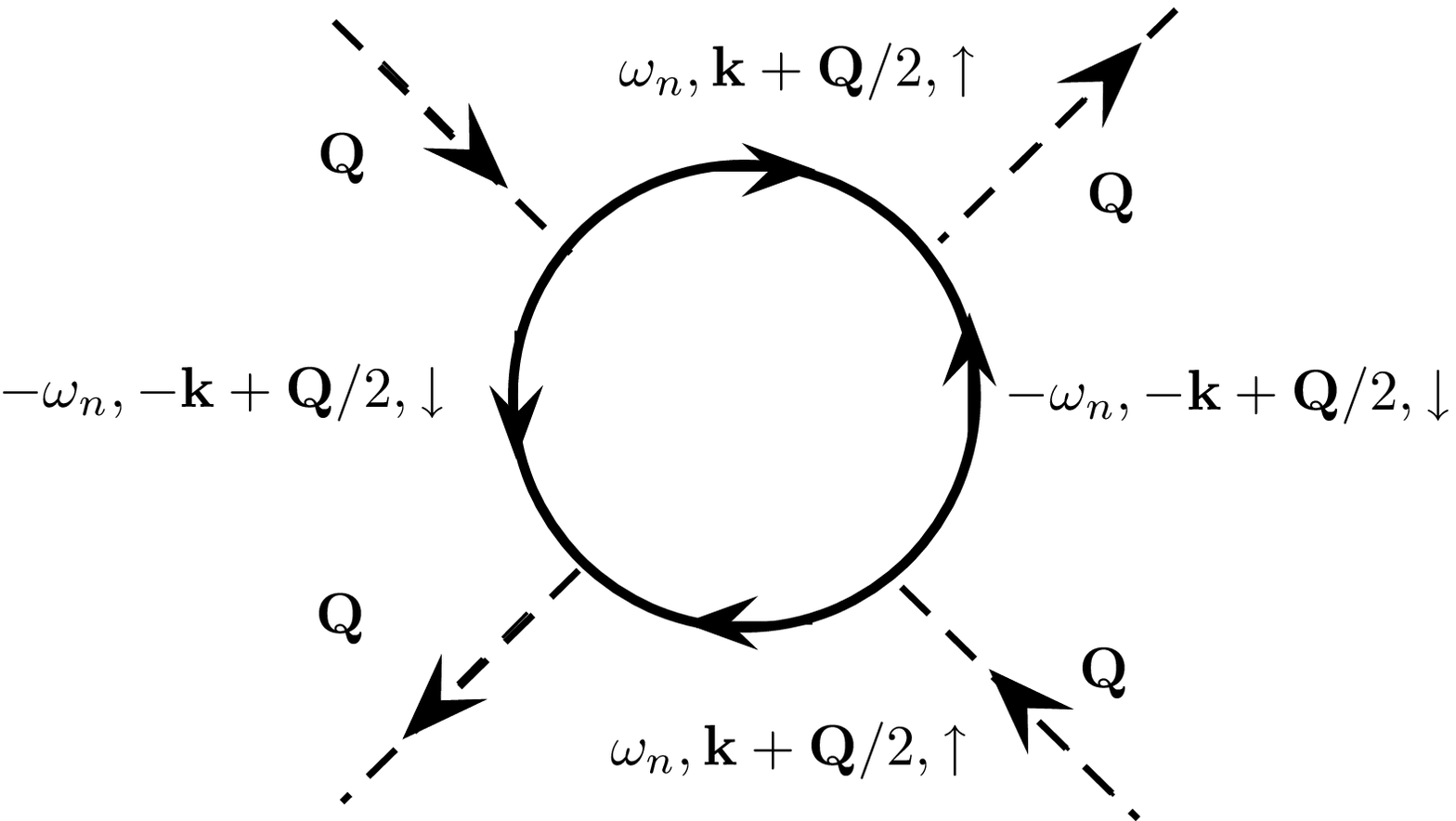}}
\subfigure[]{\includegraphics[width=0.27\textwidth]{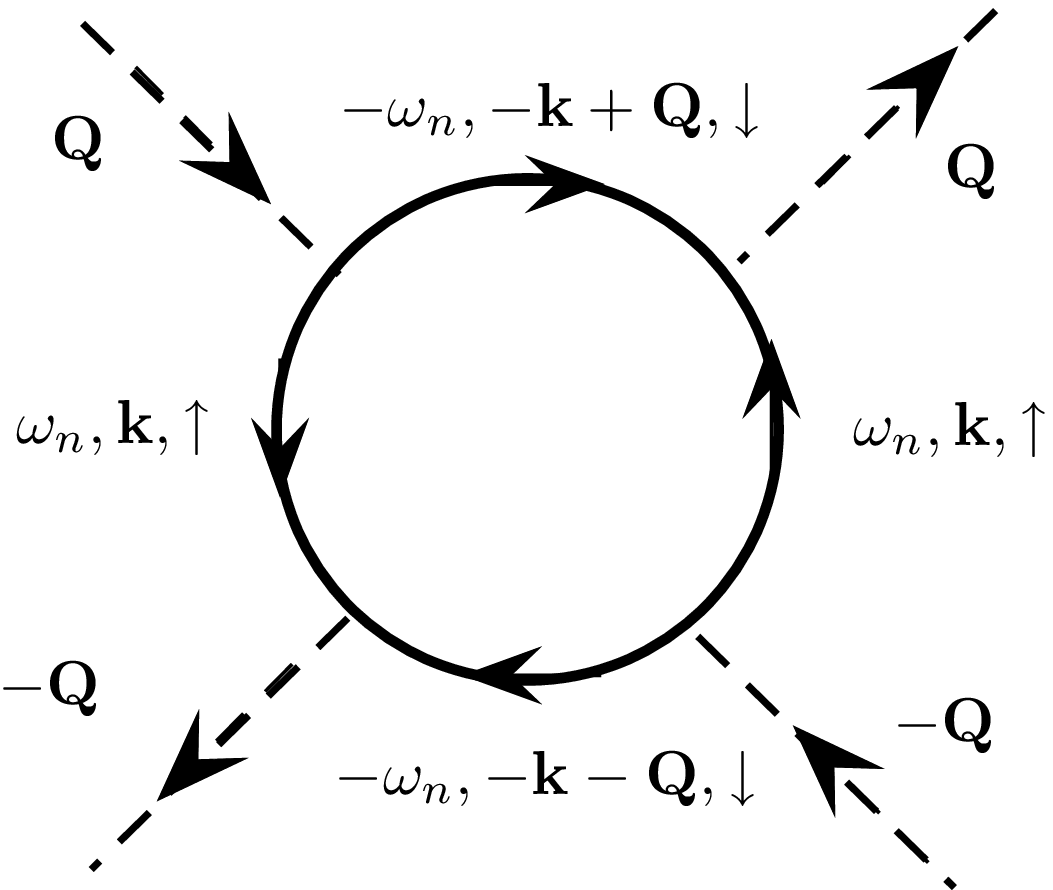}}
\end{center}
\caption{Diagrams that contribute to the coefficient $C_4^{\rm PDW}$.}
\label{fig:c4PDW}
\end{figure}
The algebraic expression for the diagram in Fig. \ref{fig:c4PDW}a is given by Eq. \eqref{c4FF}. We need only to compute the diagram in 
Fig. \ref{fig:c4PDW}b. 
The expression for this diagram is given by:
\begin{align}
\begin{split}
I_2=&N(E_F)\int_{0}^{2\pi}d\theta\frac{\cos^4 2\theta}{8\pi^2TQ/2\cos(\theta-\phi)}\left\{\text{Im}\left[\psi ^{(1)}
\left(\frac12-i\frac{\delta}{2\pi T}\cos2\theta-i\frac{Q/2}{2\pi T}\cos(\theta-\phi)\right)\right]\right.\\
&\left.\qquad\qquad\qquad\qquad\qquad\qquad\qquad\quad-\text{Im}\left[
\psi ^{(1)}\left(\frac12-i\frac{\delta}{2\pi T}\cos2\theta+i\frac{Q/2}{2\pi T}\cos(\theta-\phi)\right)\right]\right\}
\end{split}
\end{align}
We can write then:
\begin{equation}
 \frac{c^{\text{PDW}}_4}{N(E_F)}=2\frac{c^{\text{FF}}_4}{N(E_F)}+4I_2
\end{equation}
where the factor of two in the first term comes from ${\bm Q}\rightarrow -{\bm Q}$ in the diagram in Fig. \ref{fig:c4PDW}(a) and the factor of 
4 in the second term is computed in a similar way.
\end{enumerate}

\item Bidirectional PDW state

\begin{enumerate}[label=\emph{\roman*})] 
\item Quadratic term $c^{\text{Bi}}_2$.
\begin{figure}[hbt]
\begin{center}
\includegraphics[width=0.6\textwidth]{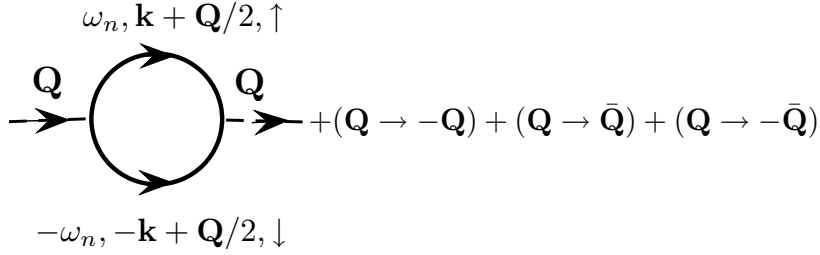}
\end{center}
\caption{Susceptibility for bidirectional superconducting order.}
\label{fig:bidirectional-LO-SC-susceptibility-diagram}
\end{figure}

For the quadratic term we need to compute $\displaystyle\frac{1}{2!}\langle S^2_{\text{int}}\rangle_0$, which can be represented by the 
diagram of Fig. \ref{fig:bidirectional-LO-SC-susceptibility-diagram}, where $\bar{{\bm Q}}=R_{\pi/2}{\bm Q}$ is the wave vector $\bm Q$ 
rotated by $\pi/2$. Each term yields the same contribution.

 Now remember that the bidirectional state have four plane waves with wave vectors 
${\bm Q}$, $-{\bm Q}$, ${\bar{\bm Q}}$, $-{\bar{\bm Q}}$ so $\sum_{{\bm q}}\frac{|\Delta_{{\bm q}}|^2}{g}=4\frac{|\Delta_{{\bm Q}}|^2}{g}$ where we used that, in order to minimize the fee energy, 
$|\Delta_{{\bm Q}}|=|\Delta_{-{\bm Q}}|=|\Delta_{{\bar{\bm Q}}}|=|\Delta_{-{ \bar{\bm Q}}}|$. Therefore we find
\begin{equation}
\frac{c^{\text{Bi}}_2}{N(E_F)}=4\frac{c^{\text{FF}}_2}{N(E_F)}
\label{c2Bi}
\end{equation}
where $T$, $Q$ and $\delta$ are in units of the BCS gap $\Delta_{BCS}$

\item Quartic term $c^{\text{Bi}}_4$.

\begin{figure}[hbt]
\begin{center}
\subfigure[]{\includegraphics[width=0.3\textwidth]{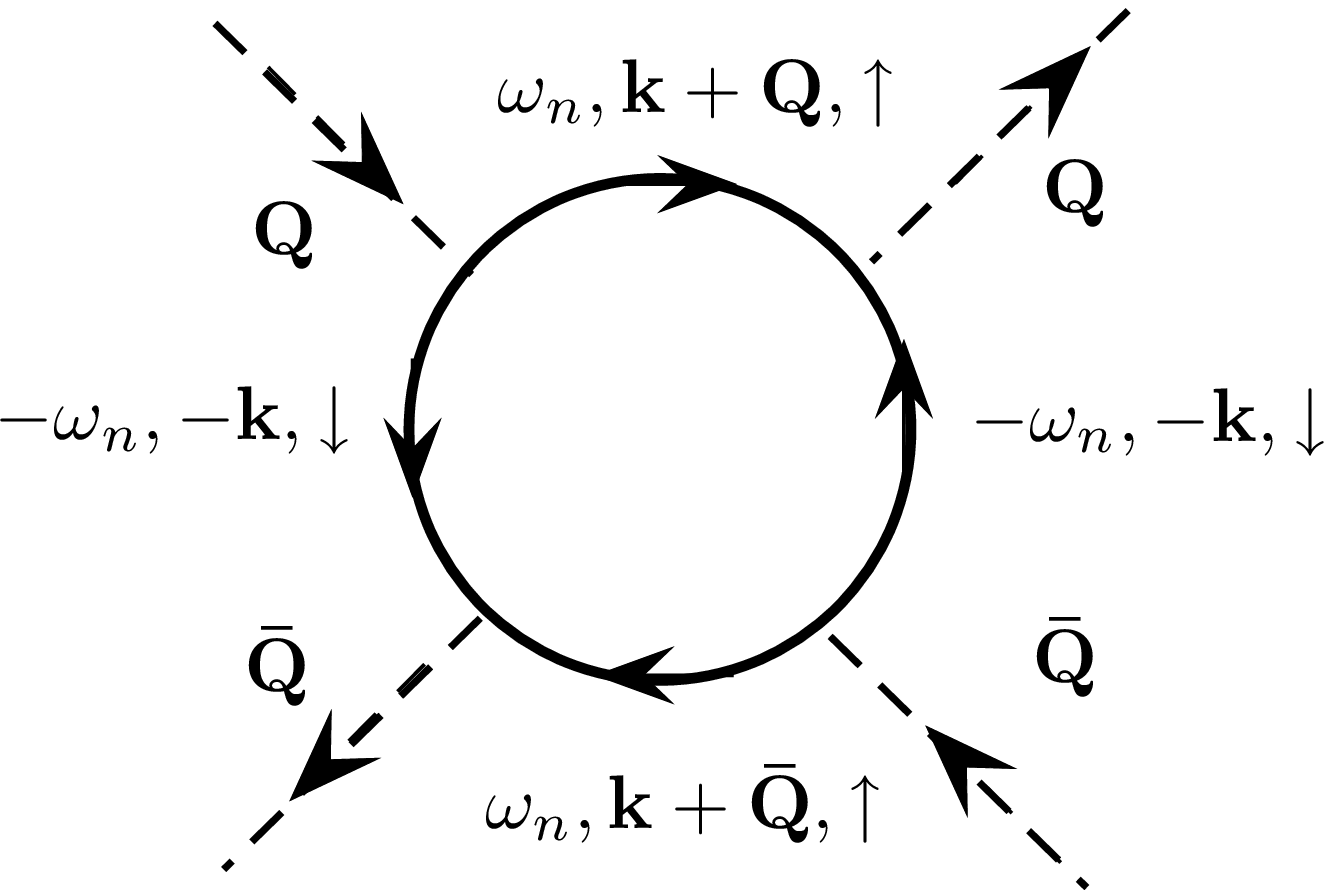}}
\subfigure[]{\includegraphics[width=0.47\textwidth]{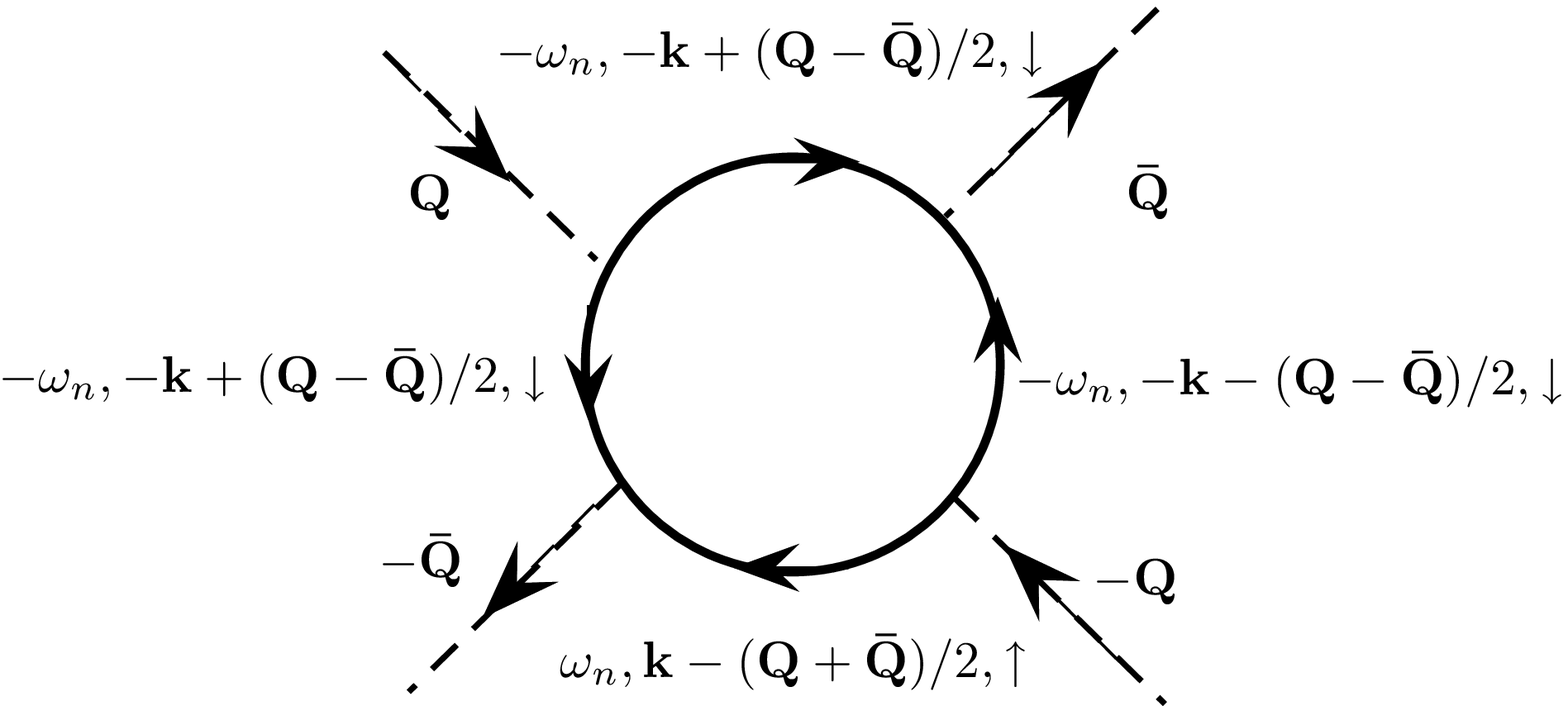}}
\end{center}
\caption{}
\label{fig:c4PDW-bi}
\end{figure}
For the quartic term we need  several diagrams. First we have contributions from the diagrams in Fig. 
\ref{fig:c4PDW}(a) and Fig. \ref{fig:c4PDW}(b), to which we must add the contributions from the ordering wave vector along the rotated direction, 
${\bm Q}\rightarrow\bar{{\bm Q}}$.

We also have the two new type of diagrams shown in Fig. \ref{fig:c4PDW-bi}(a) and Fig. \ref{fig:c4PDW-bi}(b), whose analytic expressions are
\begin{align}
\begin{split}
\frac{c_{4B1}}{N(E_F)}=&\frac{1}{4\pi QT}\int_{0}^{2\pi}\frac{d\theta}{2\pi}\frac{4\cos^4(2\theta)}{(\cos(\theta-\phi)-\sin(\theta-\phi))}
\left\{\text{Im}\left[\psi ^{(1)}
\left(\frac12-i\frac{\delta}{2\pi T}\cos2\theta+i\frac{Q/2}{2\pi T}\sin(\theta-\phi)\right)\right]\right.\\
&\left.\qquad\qquad\qquad\qquad\qquad\qquad\qquad\quad-\text{Im}\left[
\psi ^{(1)}\left(\frac12-i\frac{\delta}{2\pi T}\cos2\theta+i\frac{Q/2}{2\pi T}\cos(\theta-\phi)\right)\right]\right\}\\
\frac{c_{4B2}}{N(E_F)}=&\int_{0}^{2\pi}\frac{d\theta}{2\pi}\frac{4\cos^4(2\theta)}{Q^2(\sin^2(\theta-\phi)-\cos^2(\theta-\phi))}
\text{Re}\left\{\psi\left(\frac12+i\frac{\delta}{2\pi T}\cos2\theta+i\frac{Q/2}{2\pi T}\sin(\theta-\phi)\right)\right.\\
&\left.\qquad\qquad\qquad\qquad\qquad\qquad\qquad\qquad\qquad\quad
+\psi\left(\frac12+i\frac{\delta}{2\pi T}\cos2\theta-i\frac{Q/2}{2\pi T}\sin(\theta-\phi)\right)\right.\\
&\left.\qquad\qquad\qquad\qquad\qquad\qquad\qquad\qquad\qquad\quad
-\psi\left(\frac12+i\frac{\delta}{2\pi T}\cos2\theta+i\frac{Q/2}{2\pi T}\cos(\theta-\phi)\right)\right.\\
&\left.\qquad\qquad\qquad\qquad\qquad\qquad\qquad\qquad\qquad\quad
-\psi\left(\frac12+i\frac{\delta}{2\pi T}\cos2\theta-i\frac{Q/2}{2\pi T}\cos(\theta-\phi)\right)\right\}
\end{split}
\end{align}
Let us mention that in the diagram shown in Fig. \ref{fig:c4PDW-bi}(a) we could take $\vec{Q}\to -\vec{Q}$, $\bar{{\bm Q}}\to -\bar{{\bm Q}}$, and
$\vec{Q}\to -\vec{Q}$ and $\bar{{\bm Q}}\to -\bar{{\bm Q}}$. Each of these three extra diagrams will give
exactly the same contribution that $c_{4B1}$, hence forth the factor of 4 in the term $c_{4B1}$ in Eq. \ref{c4bi}. 

We can finally write:
\begin{equation}
 \frac{c^{\text{Bi}}_4}{N(E_F)}=\frac{4}{N(E_F)}(c^{\text{FF}}_4+2I_2+4c_{4B1}+2c_{4B2})
 \label{c4bi}
\end{equation}
\end{enumerate}
\item double helix (2H) state.

\begin{enumerate}[label=\emph{\roman*})] 
\item Quadratic term $c^{\text{2H}}_2$.\\
This is just given by:
\begin{equation}
\frac{c^{\text{2H}}_2}{N(E_F)}=2\frac{c^{\text{FF}}_2}{N(E_F)}
\label{c22H}
\end{equation}
where as usual $T$, $Q$ and $\delta$ are in units of $\Delta_{BCS}$

\item Quartic term $c^{\text{2H}}_4$.\\
For the quadratic term we have, in addition to the diagram in Fig. \ref{fig:c4FF-diagram} and the same diagram
taking ${\bm Q}\to\bar{{\bm Q}}$, the diagram shown in \ref{fig:c4PDW-bi}(a).
We can write then:
\begin{equation}
 \frac{c^{\text{2H}}_4}{N(E_F)}=2\frac{c^{\text{FF}}_4}{N(E_F)}+4c_{4B1}
\end{equation}
\end{enumerate}

\item Time-reversal breaking bidirectional PDW state.
The only change between this state and the bidirectional PDW state, is that the $c_{4B2}$ term in $c^{\text{Bi2}}_4$ enters
with a minus sign. So:
\begin{align}
\frac{c^{\text{Bi2}}_2}{N(E_F)}&=4\frac{c^{\text{FF}}_2}{N(E_F)}\\
 \frac{c^{\text{Bi2}}_4}{N(E_F)}&=\frac{4}{N(E_F)}(c^{\text{FF}}_4+2I_2+4c_{4B1}-2c_{4B2})
 \label{c4Bi2}
\end{align}

\end{enumerate}

Now that we have computed the coefficients for all the SC states we need to see which one has less energy. Using that 
$F_{\text{min}}=\displaystyle-\frac{c_2^2}{4c_4}$, we have that:
\begin{align}
 F^{\text{FF}}_{\text{min}}=&-\frac{(c^{\text{FF}}_2)^2}{4c^{\text{FF}}_4}=-\frac{(c^{\text{FF}}_2)^2}{4}\frac{1}{c^{\text{FF}}_4}\\
 F^{\text{PDW}}_{\text{min}}=&-\frac{(c^{\text{PDW}}_2)^2}{4c^{\text{PDW}}_4}=-\frac{(2c^{\text{FF}}_2)^2}{4(2c^{\text{FF}}_4+4I_2)}
=-\frac{(c^{\text{FF}}_2)^2}{4}\frac{2}{(c^{\text{FF}}_4+2I_2)}\\
 F^{\text{Bi}}_{\text{min}}=&-\frac{(c^{\text{Bi}}_2)^2}{4c^{\text{Bi}}_4}=-\frac{(4c^{\text{FF}}_2)^2}{4\cdot4(c^{\text{FF}}_4+2I_2+4c_{4B1}+2c_{4B2})}
=-\frac{(c^{\text{FF}}_2)^2}{4}\frac{4}{(c^{\text{FF}}_4+2I_2+4c_{4B1}+2c_{4B2})}\\
F^{\text{2H}}_{\text{min}}=&-\frac{(c^{\text{2H}}_2)^2}{4c^{\text{2H}}_4}=-\frac{(2c^{\text{FF}}_2)^2}{4(2c^{\text{FF}}_4+4c_{4B1})}
=-\frac{(c^{\text{FF}}_2)^2}{4}\frac{2}{(c^{\text{FF}}_4+2c_{4B1})}\\
F^{\text{Bi2}}_{\text{min}}=&-\frac{(c^{\text{Bi2}}_2)^2}{4c^{\text{---}}_4}=-\frac{(4c^{\text{FF}}_2)^2}{4\cdot4(c^{\text{FF}}_4+2I_2+4c_{4B1}-2c_{4B2})}
=-\frac{(c^{\text{FF}}_2)^2}{4}\frac{4}{(c^{\text{FF}}_4+2I_2+4c_{4B1}-2c_{4B2})}
\end{align}
In order to see which state has less energy we need to compute numerically $\displaystyle \frac{1}{c^{\text{FF}}_4}$, 
$\displaystyle \frac{2}{(c^{\text{FF}}_4+2I_2)}$, $\displaystyle \frac{4}{(c^{\text{FF}}_4+2I_2+4c_{4B1}+2c_{4B2})}$.
$\displaystyle\frac{2}{(c^{\text{FF}}_4+2c_{4B1})}$ and $\displaystyle\frac{4}{(c^{\text{FF}}_4+2I_2+4c_{4B1}-2c_{4B2})}$
and see which of these terms is larger.
As it was mentioned above, we found that the unidirectional and bidirectional PDW states have less energy than the time-reversal breaking
SC states (FF, 2H and Bi2). 
In addition, we found that for $T\gtrsim 0.2$ the bidirectional PDW state has less energy than the unidirectional PDW state.

\end{widetext}

\end{document}